\documentclass[12pt, onecolumn, oneside, draftclsnofoot]{IEEEtran}
\usepackage{ amssymb, graphicx, dsfont, setspace, url}

\usepackage[cmex10]{amsmath}
\usepackage{amsmath, amssymb, graphicx, setspace,url, amsfonts, amsthm, multirow,stmaryrd,mathabx,MnSymbol}
\usepackage{graphicx,caption,subcaption, dsfont}

\newtheorem{lemma}{Lemma}
\newtheorem{remark}{Remark}
\newtheorem{defin}{Definition}
\newtheorem{prop}{Proposition}

\newtheorem{example}{Example}
\newtheorem{theorem}{Theorem}

\usepackage[T1]{fontenc}
\usepackage[latin9]{inputenc}

\usepackage{setspace}

\begin{document}
\setcounter{page}{1}

\title{Code Construction and Decoding Algorithms for Semi-Quantitative Group Testing with Nonuniform Thresholds} 
\author{Amin~Emad and Olgica Milenkovic\\
\today
\thanks{This work was presented in part at the IEEE 2014 International Symposium on Information Theory (ISIT'14)~\cite{EM14}.}
\thanks{The authors are with the Department of Electrical and Computer Engineering, University of Illinois at Urbana-Champaign, Urbana, IL. (e-mail: emad2@illinois.edu; milenkov@illinois.edu).}
}

\maketitle
\thispagestyle{empty}
\maketitle

\begin{abstract}
We analyze a new group testing scheme, termed semi-quantitative group testing, which may be viewed as a concatenation of an adder channel and a discrete quantizer. Our focus is on non-uniform quantizers with arbitrary thresholds. For the most general semi-quantitative group testing model, we define three new families of sequences capturing the constraints on the code design imposed by the choice of the thresholds. The sequences represent extensions and generalizations of $B_h$ and certain types of super-increasing and lexicographically ordered sequences, and they lead to code structures amenable for efficient recursive decoding. We describe the decoding methods and provide an accompanying computational complexity and performance analysis. 
\end{abstract}

\section{Introduction}
Group testing is a family of pooling methods designed to efficiently identify relatively small subsets of subjects with some particular characteristic within a large collection of elements~\cite{DH06}. Rather than testing each subject individually, subgroups of subjects are tested simultaneously. The low abundance of the subjects of interest allows for determining their exact identity with a small number of tests compared to the number of test elements. Given the ubiquitous nature of the questions it addresses, this classical group testing paradigm has found many applications in communication theory, signal processing, computer science, and computational biology~\cite{W85}-\cite{DSMB09}.

A number of extensions of classical group testing (CGT) models have also been considered in the literature~\cite{DH06,DH00},~\cite{D43}-\cite{D04}, including threshold group testing (TGT)~\cite{D06} and quantitative group testing (QGT)~\cite{DH00},~\cite{D04}. In the CGT model~\cite{D43}, the result of a test equals $0$ if the test does not include subjects of interest (i.e., ``defectives''), and $1$ otherwise. In the TGT model, if the number of defectives in a test is smaller than a lower threshold, the test outcome equals $0$; if the number of defectives is larger than an upper threshold, the test outcome equals $1$; and if the number of defectives is between the lower and upper threshold, the test result is arbitrary, either equal to $0$ or $1$. 
In QGT, the result of a test equals the exact number of defectives appearing in the test. 

Group testing (GT) is closely related to the field of compressed sensing (CS)~\cite{CRT06,DO06}, and in particular, integer compressed sensing~\cite{DMICS}, in so far that both group testing and compressed sensing seek to recover a sparse unknown vector through a small set of measurements. The CS model particularly shares a number of features with the QGT model, since in both of these problems the vector of measurements is obtained through the product of a sensing matrix (or test matrix) with the unknown sparse vector. However, due to the limited precision in obtaining the measurements, the linearity assumption of the measurements does not apply in many practical applications. In \cite{DPM09,DM09-Long,ZBC10}, quantized compressed sensing (QCS) was introduced to overcome the limitation of infinite precision in CS, while in~\cite{AM11,AM13}, we introduced the semi-quantitative group testing (SQGT) paradigm to overcome this issue in the the GT framework. The motivation for using quantized linear measurements in GT stems from applications in genotyping~\cite{AM13} and conflict resolution in multiple access channel (MAC) communication. Since a detailed description of these applications and the motivation behind SQGT is provided in~\cite{AM13}, we refer the interested reader to this publication for more details. 

In SQGT, the result of a test is a value from a non-binary alphabet that depends on the number of defectives through a fixed set of thresholds. Simply put, an SQGT model represents a concatenation of a QGT model and a quantizer which models the limited precision in obtaining the measurements. In \emph{nonadaptive} SQGT, each subject is assigned a unique binary or non-binary vector (codeword) of length equal to the total number of tests. We always assume that the available alphabet size for constructing the codewords is a fixed, finite integer which is imposed by the application of interest. It is customary to arrange the codewords as columns of a matrix, subsequently referred to as the test matrix (codebook). Each coordinate in the codeword assigned to a subject corresponds to a test, and its value reflects the ``strength'' of the subject in the test~\cite{AM11},~\cite{AM13}. The interpretation of the word ``strength'' depends on the application at hand: for example, ``strength'' may correspond to the power level of a MAC user, or, it may correspond to the concentration of the genetic material of an individual. Two important families of SQGT codes, SQ-disjunct and SQ-separable, were introduced and analyzed in our companion papers~\cite{AM11},~\cite{AM13}. In the same work, constructions for uniformly quantized SQGT codes were presented, based on number-theoretic sequence selection methods. 

Although the motivation behind SQGT and QCS is the limited precision in obtaining the measurements, there are some major differences between these models. In QCS, the entries of the sensing matrix are real or complex numbers, while in SQGT the entries of the test matrix are positive integers. Integer-valued test matrices are used in applications where the subjects to be tested come as a whole (or multiples of a predetermined fixed value) and cannot be ``subdivided'' into \emph{real-valued} parts. For example, in the coin-weighing problem, if one has $n$ bags of coins, where each bag contains $q-1$ identical coins, and some of the bags have counterfeit coins, one can use tests of alphabet size $q$ to find the bag containing counterfeit coins, with many fewer experiments than using binary tests. Another application of integer-valued test matrices is in applications where there is more flexibility in choosing the alphabet size of the test matrix (as compared to binary alphabet test matrices), but a real-valued alphabet may not be practical due to ``limited precision''. Yet another application of such matrices is in scenarios where some robustness to errors and noise is needed in the testing schemes; integers, unlike reals, are spaced discretely, which ensures a form of error protection in forming the test matrix.

Another difference between QCS and SQGT is that in the latter case, one is mostly interested in constructing test matrices that are capable of identifying the defectives with \emph{zero error probability} in the presence of errors; however, in QCS, a small error value is tolerated and only an approximation of the sparse unknown vector is sought. As a result, in prior work on QCS~\cite{DPM09,ZBC10}, the focus has been on developing algorithms and deriving distortion-rate functions and error estimates for the obtained solutions. One should note that the reason zero error probability may be achieved in SQGT is that the test matrix is integer-valued as opposed to real-valued, which allows us to construct robust test matrices capable of tolerating errors in the vector of test results\footnote{Another line of research on this subject relates to ``small-error'' information-theoretic limits of GT, akin to the work described in~\cite{Jaggi2014,Jaggi2015}. There, the goal is to derive algorithms that succeed "with high probability" rather than with probability one.}. 

The central theme of this work is non-adaptive SQGT with non-uniformly spaced thresholds, the most general framework in which one can study this testing scheme. Although many special choices of GT with thresholds such as CGT, TGT, and QGT have been studied in the literature, there is not much known regarding construction of code matrices for SQGT with arbitrary thresholds. The only exception are some of the preliminary results derived in our companion paper~\cite{AM13}. 

The contributions of this paper are threefold. First, we describe three new families of integer sequences with properties that can be utilized in SQGT. Second, we describe constructions for SQGT test matrices using these new sequences and show that the resulting schemes are capable of identifying the defectives in SQGT with arbitrary thresholds. Third, for each of these constructions we describe a computationally efficient decoding algorithm that can identify the defectives with zero error probability in the presence of errors in the vector of test results. 

The paper is organized as follows. In Section~\ref{sec:model} we introduce the SQGT model and describe the relevant terminology. In Section~\ref{sec:disjunct} we describe the notion of SQ-separable test matrices and provide a summary of our results. The derivations of our main results are presented in Sections~\ref{sec:bh},~\ref{sec:sqlos} and~\ref{sec:sqlol}. In Section~\ref{sec:bh}, we introduce \emph{quantized} $B_h$ sequences, and describe how to use their elements in conjunction with binary disjunct codes to construct separable SQGT codes. There, we also describe construction methods for quantized $B_h$ sequences as well as decoding methods for the resulting codes. In Sections~\ref{sec:sqlos} and~\ref{sec:sqlol}, we introduce the notion of semi-quantitative lexicographical orders and their corresponding sequences and describe how to construct them. In addition, we describe constructions for SQ-separable matrices using these sequences and present two computationally efficient decoding algorithms for these matrices.

\section{The Semi-quantitative Group Testing Model}\label{sec:model}

Throughout the paper, we use bold-face upper-case and bold-face lower-case letters to denote matrices and vectors, respectively. Calligraphic letters are reserved for sets and sequences. In addition, asymptotic notations such as $o(\cdot)$ and $O(\cdot)$ are used in a standard manner. For a parameter $g$, we use the notation $O_g(\cdot)$ to mean that the constant factor in this asymptotic notation is a function of the parameter $g$. 

Let $\mathbb{Z}^+$ denote the set of positive integers. For an integer $n\in\mathbb{Z}^+$, we write $[n]:=\{0,1,\dots,n-1\}$ and $\llbracket{n}\rrbracket :=\{1,2,\dots,n\}$. With slight abuse of notation, we use $\mathcal{A}\!=\!\{\!\alpha_1,\alpha_2,\dots,\alpha_K\!\}$ to denote both a set and/or a sequence consisting of $K$ positive integers. The exact meaning will be apparent from the context, and it will depend on which property of $\mathcal{A}$ is being discussed. Note that for a set of positive integers $\mathcal{A}$, one can view the natural ordering of the elements of $\mathcal{A}$ as the corresponding sequence.

Let $n$, $m$, and $d$ denote the number of test subjects, the number of tests, and the number of defectives, respectively. With each subject, we associate a unique $q$-ary vector, $q \geq 2$, of length $m$, termed a codeword. Due to the one-to-one correspondence between the codewords and test subjects, with some abuse of notation we use $\mathcal{D}$ to denote both the set of defectives and the set of codewords assigned to the defectives. Each coordinate of a codeword corresponds to a test. If $\mathbf{x}_i\in[q]^m$ denotes the codeword of the $i^{\textnormal{th}}$ subject, then the $k^{\textnormal{th}}$ coordinate of $\mathbf{x}_i$, denoted by $\mathbf{x}_i(k)$, represents the ``strength'' of the $i^{\textnormal{th}}$ subject in the $k^{\textnormal{th}}$ test.
The set of codewords is represented by the codebook $\mathbf{C}\in[q]^{m\times n}$. 

The result of SQGT tests can be represented as a vector $\mathbf{y}\in{[Q]}^m$, called the vector of test results. Each test outcome depends on the number of defectives $d$ and their strengths through a quantization function $f_{\boldsymbol{\eta}}(\cdot)$, defined as follows. 

\begin{defin}\label{def_qua}
For a set of thresholds $\boldsymbol{\eta}=[\eta_0=0,\eta_1,\dots,\eta_{Q}]^T$ and a scalar $\alpha\in{\mathbb{Z}^{+}}$, we define the quantization function $f_{\boldsymbol{\eta}}:{\mathbb{Z}^{+}}\mapsto [Q]$ as
\begin{align}\nonumber
f_{\boldsymbol{\eta}}(\alpha)=r\  \  \  \  \  \text{if}\  \  \  \  \  \eta_r\leq \alpha<\eta_{r+1},
\end{align}
where $r\in[Q]$. In words, the function $f_{\boldsymbol{\eta}}(\alpha)$ returns the index of the quantization bin that contains its argument.
\end{defin}
For a vector of positive integers $\boldsymbol{\alpha}$, $f_{\boldsymbol{\eta}}(\boldsymbol{\alpha})$ is a vector with each entry equal to the quantization of the corresponding entry of $\boldsymbol{\alpha}$ according to Def.~\ref{def_qua}.
For two scalars $\alpha,\alpha'\in{\mathbb{Z}^{+}}$, and a set of thresholds $\boldsymbol{\eta}$, we write $\alpha\succ_{\boldsymbol{\eta}}\alpha'$ to indicate that $f_{\boldsymbol{\eta}}(\alpha)>f_{\boldsymbol{\eta}}(\alpha')$. Next, we define the syndrome of a set of codewords using $f_{\boldsymbol{\eta}}(\cdot)$.

\begin{defin}[\textbf{Syndrome of a set of codewords}]\label{def_syndrome}
Let $\mathcal{X}=\{\mathbf{x}_1,\mathbf{x}_2,\dots,\mathbf{x}_s\}=\{\mathbf{x}_j\}_{1}^s$ be a set of $s\geq 1$ codewords of length $m$ in a SQGT model with thresholds $\boldsymbol{\eta}=[\eta_0=0,\eta_1,\eta_2,\dots,\eta_Q]^T$. The syndrome of  $\mathcal{X}$, denoted by $\mathbf{y}_{\!_\mathcal{X}}\in{[Q]}^m$, is defined as $\mathbf{y}_{\!_\mathcal{X}}=f_{\boldsymbol{\eta}}\left(\sum_{j=1}^{s}\mathbf{x}_{j}\right)$.
\end{defin}

\begin{figure}
\includegraphics[width=\textwidth]{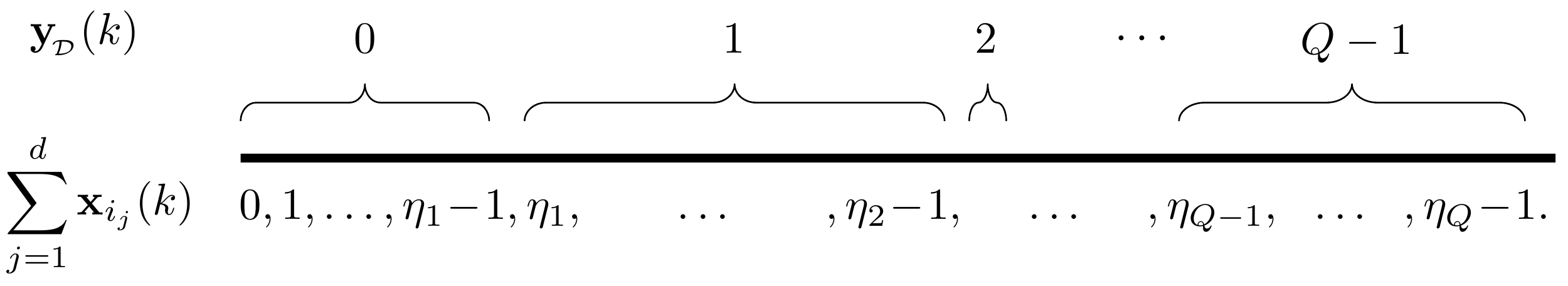}
\centering
\vspace{-5pt}
\caption{The outcome of the $k^{\text{th}}$ test (in the absence of error) as a function of $\sum_{j=1}^d \mathbf{x}_{i_j}(k)$.}
\label{fig:test-result}
\end{figure}

By this definition, in the absence of any errors, the vector of test results is equal to the syndrome of defectives, i.e. $\mathbf{y}=\mathbf{y}_{\!_\mathcal{D}}$. However, when errors occur, some entries of $\mathbf{y}$ may differ from $\mathbf{y}_{\!_\mathcal{D}}$. In particular, if $e$ tests are erroneous, we assume that $e$ entries of $\mathbf{y}_{\!_\mathcal{D}}$ have changed to an arbitrary value in $[Q]$\footnote{Note that this assumption corresponds to the case in which no information is available regarding the pattern of errors (i.e. worst case scenario). However, more informative assumptions regarding the error pattern can be considered to simplify the problem, e.g. errors that change the outcome of a test to the value corresponding to an adjacent bin.}. The relationship between the syndrome of defectives and the strength of the defectives in a test is illustrated in Fig.~\ref{fig:test-result}. One should note that an underlying assumption in the SQGT model is that $\eta_Q>d(q-1)$, which is needed to ensure that the sum of entries corresponding to defectives is always smaller than $\eta_Q$. The previously described parameters and their definitions are provided in Table~\ref{tab:notation}

Note that SQGT includes different group testing models as special cases. For example if $q=Q=2$, $\eta_1=1$, and $\eta_2=+\infty$ the SQGT model reduces to CGT. Furthermore, if $Q-1=d(q-1)$ and $\forall r\in[Q]$, $\eta_r=r$, then SQGT reduces to the quantitative (adder) model (QGT) with a possibly non-binary test matrix. 

\begin{table}[t!]\centering
\caption{Table of symbols and their definitions}
\begin{tabular}{|c|c|}
			\hline 
			Symbol & Definition \\ 
			\hline\hline
		
			$n$ & Total number of subjects \\	\hline
			$m$ & Number of tests \\				\hline
			$d$ & Number of defectives \\	\hline				
			$Q$ & Size of the output alphabet\\	\hline
			$q$	& Size of the test matrix alphabet \\ \hline	
			$\eta_l$ & The $l^{\textnormal{th}}$ threshold where $l\in\llbracket Q\rrbracket$ \\	\hline
			$\mathcal{D}$ & Set of defectives \\				\hline		
			$\mathbf{y}\in[Q]^m$ & Vector of test results \\	\hline										
			$\mathbf{C}\in[q]^{m\times n}$ & Code (test matrix) \\\hline
			$e$ & Number of errors in $\mathbf{y}$ that $\mathbf{C}$ can correct \\				\hline
			
			\end{tabular}
			\label{tab:notation}

\end{table}

\section{Superimposed Codes for SQGT and Summary of the Results}\label{sec:disjunct}
In~\cite{AM11} and~\cite{AM13}, we introduced SQ-separable test matrices to identify the defectives with zero error-probability. A $[q;Q;\boldsymbol{\eta};(l\!:\!u);e]$-SQ-separable code is a $q$-ary matrix for a SQGT model with thresholds $\boldsymbol{\eta}=[0,\eta_1,\eta_2,\dots,\eta_Q]^T$, capable of uniquely identifying a number of defectives between $l$ and $u$, $l\leq d\leq u,$ with zero error probability using a $Q$-ary vector of test results that contains up to $e$ errors. SQ-separable matrices are defined as follows.

\begin{defin}[\textbf{SQ-separable codes} \cite{AM13}]\label{SQsep}
A $m\times n$ matrix is called a $[q;Q;\boldsymbol{\eta};(l\!:\!u);e]$-SQ-separable code if for any two distinct sets of codewords (i.e. columns), $\mathcal{X}$ and $\mathcal{Z}$, satisfying $l\leq|\mathcal{X}|,|\mathcal{Z}|\leq u$, there exists a set of coordinates $\mathcal{R}$, satisfying $|\mathcal{R}|\geq 2e+1$, such that $\forall k\in\mathcal{R}$, $\mathbf{y}_{\!_{\mathcal{X}}}(k)\neq \mathbf{y}_{\!_{\mathcal{Z}}}(k)$, where $\mathbf{y}_{\!_{\mathcal{X}}}$ and $\mathbf{y}_{\!_{\mathcal{Z}}}$ are the syndromes of $\mathcal{X}$ and $\mathcal{Z}$, respectively (see Def.~\ref{def_syndrome}).
\end{defin}

Intuitively, SQ-separable codes impose the requirement that any collection of not fewer than $l$ and not more than $u$ items have a unique syndrome after quantization and in the presence of errors. One should note that ``being SQ-separable'' is a necessary condition for a text matrix to identify the defectives in an SQGT model with \emph{zero error probability} \cite{AM13}. In other words, a test matrix $\mathbf{C}$ can identify any number of defectives between $l$ and $u$ with zero error probability in the presence of up to $e$ errors if and only if $\mathbf{C}$ is a SQ-separable code. Due to their generality, it is not surprising that no universal \emph{computationally efficient} decoder is currently known that can identify the defectives using a SQ-separable matrix with zero error probability for an arbitrary set of thresholds. In order to overcome the issue of efficient decoding, pervious works in the literature have either relaxed the zero error probability requirement and have used approximation algorithms such as message passing on factor graphs \cite[Appendix A]{AM13}, or they have focused on special choices of the thresholds (e.g. all the previous work on CGT, TGT, QGT, or SQGT with uniform thresholds discussed in \cite{AM13}). Another approach that we used in \cite{AM13} was to impose extra structure on the test matrix which lead to the introduction of SQ-disjunct matrices that are endowed with simple decoders of computational complexity $O(mn)$. However, due to their strict structure, the best known SQ-disjunct matrices are not capable of fully utilizing all the information in the vector of test results to reduce the number of tests. 

In this paper, for the first time, we will introduce test matrix constructions for SQ-separable matrices that are endowed with computationally efficient zero-error decoding algorithms. These matrices use SQ-disjunct matrices of size $m_b\times n_b$ as building blocks, and depending on the available alphabet, are able to increase the number of columns (i.e. subjects) $K$-fold for a fixed number of rows (i.e. tests); in other words, for a given $m_b\times n_b$ SQ-disjunct matrix, one can construct $m_b\times Kn_b$ SQ-separable matrices, where $K$ depends on $q$, $\boldsymbol{\eta}$, and the specific construction method. More importantly, we do not focus on any special choice of thresholds, and the constructions can apply to a wide range of choices for $\boldsymbol{\eta}$. To achieve this goal, we introduce three families of integer sequences that lend themselves to SQGT code design, termed ``quantized $B_h$'',  ``type-s semi-quantitative lexicographically ordered sequences (SQLO$_s$)'' and ``type-l semi-quantitative lexicographically ordered sequences (SQLO$_l$)'' . While SQLO$_s$ and SQLO$_l$ sequences are special cases of quantized $B_h$ sequences, they exhibit a special nested structure that allows for computationally efficient decoding algorithms. These results are summarized in Table~\ref{table:summary}. Note that the aforementioned sequences have different densities, and therefore for a fixed alphabet size $q$, the largest possivle value of $K$ may be different for each sequence. As a result, in the table we used $K_b$, $K_s$, and $K_l$ to distinguish between the value of $K$ corresponding to each sequence. For a fixed choice of parameters, $K_b$ is in general larger than $K_s$ and $K_l$. This implies that a code constructed using a quantized $B_h$ sequence will require a smaller number of tests compared to the codes constructed using SQLO sequences; however, the computational complexity of its corresponding decoder will be higher as well. Relevant bounds and detailed discussions of the properties of the resulting codes are described in the following sections.

\begin{table}[t!]\centering
 \caption{A comparative summary of SQGT matrices described in this paper}
{\renewcommand{\arraystretch}{1.5}{\footnotesize\begin{tabular}{|l|c|c|c|c|c|}
			\hline 
			\textbf{Test Matrix}  & Theorem~\ref{const_Bh} &Theorem~\ref{const_SQLO} & Theorem~\ref{const_SQLO2} \\
			\hline\hline
		
			\textbf{Parameters}  & $[q;Q;\boldsymbol{\eta};(1\!:\!d);e]$ & $[q;Q;\boldsymbol{\eta};(1\!:\!d);e]$ & $[q;Q;\boldsymbol{\eta};(1\!:\!d);e]$\\[0pt]\hline
			\textbf{Type}  & SQ-separable & SQ-separable & SQ-separable \\\hline
			\textbf{Thresholds} & Arbitrary & Arbitrary & Arbitrary\\\hline			
			\textbf{Construction} & Explicit & Explicit & Explicit \\\hline
			\textbf{Num. Tests} & $O_e\!\left(d^2\log_2 \frac{n}{dK_b}\right)$&$O_e\!\left(d^2\log_2 \frac{n}{dK_s}\right)$ &$O_e\!\left(d^2\log_2 \frac{n}{dK_l}\right)$ \\\hline
			 \textbf{Features}  & Decoder of complexity & Decoder of complexity &Decoder of complexity\\
				                       &$O(\frac{mn}{K_b}+2^{K_b}(K_b+md))$, & $O(\frac{mn}{{K_s}}+dm\log m+de g_{\max}K_s)$,  & $O(\frac{mn}{K_l}+dm\log m+de g_{\max}K_l)$, \\
						            
						             & Uses $K_b$ elements of a& Uses $K_s$ elements of a & Uses $K_l$ elements of a\\
						             &quantized $B_d$ sequence &SQLO$_s(\boldsymbol{\eta},d)$ sequence &SQLO$_l(\boldsymbol{\eta},d)$ sequence\\\hline

			\end{tabular}}}
			\label{table:summary}
\end{table}

In order to gain intuition about these results, we first provide some necessary conditions for the existence of a SQ-separable matrix with parameters $[q;Q;\boldsymbol{\eta};(1\!:\!d);e]$. One necessary condition that leads to a lower bound on the alphabet size $q$ is that $q\geq\eta_1+1$. The reason for this bound is that if a SQ-separable matrix does not satisfy this inequality, all the codewords (i.e. columns of the test matrix) have the same syndrome which is an all-zero vector of length $m$. As a result, if there exists one defective, one cannot uniquely identify which codeword corresponds to the defective. In addition, one cannot distinguish between the absence of any defectives and the presence of exactly one defective. A consequence of this inequality is that if $\eta_1>1$, no binary $[2;Q;\boldsymbol{\eta};(1\!:\!d);e]$ exists for this set of thresholds, no matter what the value of other thresholds are in the SQGT model. 

Another simple necessary condition provides a lower bound on the number of tests. Since a $[q;Q;\boldsymbol{\eta};(1\!:\!d);e]$-SQ-separable code must distinguish between different sets of $d$ defectives, one must have ${n\choose d}\leq Q^m$. As a result, one must have 
\begin{align}\label{bound_SQGT1}
m\geq d\log_Q\left(\frac{n}{d}\right)(1+o(1)).
\end{align}
Although this lower bound may suggest that the results obtained in this paper, which are of the form $m=O_e\!\left(d^2\log_2 \frac{n}{dK}\right)$, are a factor of $d$ away from the best possible results, one should note that the results in this paper correspond to the most general choice of thresholds. Due to this generality of the results, they should hold for any choice of thresholds including the special case of CGT in which $\eta_1=1$ and $\eta_2=\infty$; on the other hand, it is well known that in CGT, the number of required tests to identify the defectives with zero error probability (even in the absence of any errors) satisfies~\cite{N00}
\begin{align}\label{bound_SQGT2}
 m\geq \frac{d^2}{2\log_2d}\log_2n\  (1\!+\!o(1)).
\end{align}
This implies that without imposing further constraints on the thresholds in SQGT, one cannot reduce the number of tests by an extra factor of $d$. We have shown in~\cite{AM13} that by focusing on some special cases of thresholds (e.g. equidistant thresholds), this reduction in the number of tests by a factor of $d$ is possible and one can tightly match the lower bound in~\eqref{bound_SQGT1}.

Next, we describe the idea behind the constructions introduced in this paper. The gist of our constructions is horizontal matrix concatenation, defined as follows. 
\begin{defin}[\textbf{Horizontal concatenation}]
Consider $K\geq2$ matrices $\mathbf{C}_j\in\mathbb{R}^{m\times n}$, $1\leq j\leq K$. The horizontal concatenation of these matrices is a matrix defined by $\mathbf{C}=[\mathbf{C}_1,\mathbf{C}_2\dots,\mathbf{C}_K],$ such that for $j\in\llbracket K\rrbracket$ and $l\in\llbracket n\rrbracket$, the $((j-1)n+l)^{\text{th}}$ column of $\mathbf{C}$ is equal to the $l^{\text{th}}$ column of $\mathbf{C}_j$.
\end{defin}


For the subsequently described code constructions, we use binary disjunct matrices for CGT as building blocks for constructing SQ-separable codes. For completeness, we start by defining SQ-separable codes~\cite{AM13} and binary disjunct codes for CGT~\cite{KS64},~\cite{DH06}.

.
\begin{defin}[\textbf{Binary $\boldsymbol{d}$-disjunct codes for CGT}]\label{CGTdisjunct}
A binary CGT $d$-disjunct code capable of correcting up to $e$ errors is a code of length $m$ and size $n$ with the property that for any codeword $\mathbf{z}$ and any subset of $d$ other codewords, $\mathcal{X}$, $\mathbf{z}\notin\mathcal{X}$, there exists a set of coordinates $\mathcal{R}$ of size at least $2e+1$, so that $\forall k\in\mathcal{R}$ and $\forall \mathbf{x}\in\mathcal{X}$, $\mathbf{z}(k)=1$ and $\mathbf{x}(k)=0$.
\end{defin}

Before describing the main results of this paper, we introduce a simple code construction that provides the intuition behind the derivations of the main results.

\begin{theorem}\label{const1}
Consider a SQGT system with thresholds $\boldsymbol{\eta}=[0,\eta_1,\eta_2,\eta_3,\dots,\eta_Q]^T$ where $Q \geq 4$. Fix a binary $d$-disjunct code matrix $\mathbf{C}_b$ of dimensions $m_b\times n_b$, capable of correcting up to $e$ errors. Form a matrix $\mathbf{C}$ of length $m=m_b$ and size $n=2n_b$ by concatenating $\mathbf{C}_1=\alpha_1\mathbf{C}_b$ and $\mathbf{C}_2=\alpha_2\mathbf{C}_b$ horizontally, where $\alpha_1=\eta_1$ and $\alpha_2=\max\{\eta_2,\eta_3-\eta_1\}$.  The constructed code is a $[q;Q;\boldsymbol{\eta};(1\!:\!d);e]$-SQ-separable code with $q=\max\{\eta_2,\eta_3-\eta_1\}+1$.
\end{theorem} 
\begin{IEEEproof}
Consider two distinct subsets of codewords, $\mathcal{X}_1$ and $\mathcal{X}_2$, such that $1\leq|\mathcal{X}_1|,|\mathcal{X}_2|\leq d$. Without loss of generality, assume that $|\mathcal{X}_1|\leq|\mathcal{X}_2|$. Since the two sets are distinct, $\mathcal{X}_2\backslash\mathcal{X}_1\neq\varnothing$. Let $\mathbf{z}'\in\mathcal{X}_2\backslash\mathcal{X}_1$. By construction, $\mathbf{z}'=\alpha\mathbf{z}_b$ for some $\alpha\in\{\alpha_1,\alpha_2\}$ and some binary codeword $\mathbf{z}_b$ of $\mathbf{C}_b$. Let $\mathbf{z}''$ be another codeword of $\mathbf{C}$ with the same support as $\mathbf{z}'$, obtained by multiplying $\mathbf{z}_b$ by $\{\alpha_1,\alpha_2\} \backslash \{\alpha\}$. 

If $\mathbf{z}''\notin\mathcal{X}_1$, then by the construction of $\mathbf{C}$ and Def.~\ref{CGTdisjunct}, there exists a set of coordinates $\mathcal{R}$ of size at least $2e+1$, such that $\forall k\in\mathcal{R}$, $\mathbf{z}'(k)\geq\alpha_1=\eta_1$ and $\mathbf{x}(k)=0$, $\forall \mathbf{x}\in\mathcal{X}_1$. Since $\forall k\in\mathcal{R}$, $\sum_{\mathbf{x}\in\mathcal{X}_2}\mathbf{x}(k)\geq\mathbf{z}'(k)\geq\eta_1$, and $\sum_{\mathbf{x}\in\mathcal{X}_1}\mathbf{x}(k)=0$, it follows that 
\begin{equation}\nonumber
\mathbf{y}_{\!_{\mathcal{X}_2}}(k)\geq\mathbf{y}_{\!_{\{\mathbf{z}'\}}}(k)>\mathbf{y}_{\!_{\mathcal{X}_1}}(k).
\end{equation}
On the other hand, if $\mathbf{z}''\in\mathcal{X}_1\cap\mathcal{X}_2$, there exists a set of coordinates $\mathcal{R}$ of size at least $2e+1$, such that $\forall k\in\mathcal{R}$, $\mathbf{z}'(k)\in\{\alpha_1,\alpha_2\}$, $\mathbf{z}''(k)\in\{\alpha_1,\alpha_2\}$, and $\mathbf{x}(k)=0$ $\forall\mathbf{x}\in\mathcal{X}_1\backslash \{\mathbf{z}''\}$. Since $\forall k\in\mathcal{R}$, $\sum_{\mathbf{x}\in\mathcal{X}_2}\mathbf{x}(k)\geq\mathbf{z}'(k)+\mathbf{z}''(k)=\alpha_1+\alpha_2=\max\{\eta_1+\eta_2,\eta_3\}\geq\eta_3$ and $\sum_{\mathbf{x}\in\mathcal{X}_1}\mathbf{x}(k)\leq\alpha_2<\eta_3$, and since $\eta_Q>\eta_1+\max\{\eta_2,\eta_3-\eta_1\}$, it follows that
\begin{equation}\nonumber
\mathbf{y}_{\!_{\mathcal{X}_2}}(k)\geq\mathbf{y}_{\!_{\{\mathbf{z}',\mathbf{z}''\}}}(k)>\mathbf{y}_{\!_{\mathcal{X}_1}}(k).
\end{equation}
If $\mathbf{z}''\in\mathcal{X}_1\backslash\mathcal{X}_2$, we have to separately analyze two cases: if $\mathbf{z}'={\alpha_2}\mathbf{z}_b$, then there exists a set of coordinates $\mathcal{R}$ of size at least $2e+1$, such that $\forall k\in\mathcal{R}$, $\mathbf{z}'(k)=\alpha_2$, $\mathbf{z}''(k)=\alpha_1$, and $\mathbf{x}(k)=0$ $\forall\mathbf{x}\in\mathcal{X}_1\backslash\{\mathbf{z}''\}$. Since $\forall k\in\mathcal{R}$, $\sum_{\mathbf{x}\in\mathcal{X}_2}\mathbf{x}(k)\geq\mathbf{z}'(k)=\alpha_2\geq\eta_2$, and $\sum_{\mathbf{x}\in\mathcal{X}_1}\mathbf{x}(k)=\alpha_1=\eta_1<\eta_2$, it follows that 
\begin{equation}\nonumber
\mathbf{y}_{\!_{\mathcal{X}_2}}(k)\geq\mathbf{y}_{\!_{\{\mathbf{z}'\}}}(k)>\mathbf{y}_{\!_{\mathcal{X}_1}}(k).
\end{equation}
However, for the case that $\mathbf{z}''\in\mathcal{X}_1\backslash\mathcal{X}_2$ and $\mathbf{z}'={\alpha_1}\mathbf{z}_b$, there exists a set of coordinates $\mathcal{R}$ of size at least $2e+1$, such that $\forall k\in\mathcal{R}$, $\mathbf{z}'(k)=\alpha_1$, $\mathbf{z}''(k)=\alpha_2$, and $\mathbf{x}(k)=0$ $\forall\mathbf{x}\in\mathcal{X}_2\backslash\{\mathbf{z}'\}$. Since $\forall k\in\mathcal{R}$, $\sum_{\mathbf{x}\in\mathcal{X}_1}\mathbf{x}(k)\geq\mathbf{z}''(k)=\alpha_2\geq\eta_2$, and $\sum_{\mathbf{x}\in\mathcal{X}_2}\mathbf{x}(k)=\alpha_1=\eta_1<\eta_2$, we conclude that 
\begin{equation}\nonumber
\mathbf{y}_{\!_{\mathcal{X}_2}}(k)<\mathbf{y}_{\!_{\{\mathbf{z}''\}}}(k)\leq\mathbf{y}_{\!_{\mathcal{X}_1}}(k).
\end{equation}
This completes the proof.
\end{IEEEproof}

In~\cite[Construction 1]{AM13}, it was shown that multiplying a binary $d$-disjunct code of dimension $m_b\times n_b$ by $\eta_1$ results in a SQ-disjunct code of the same dimension. On the other hand, Thm.~\ref{const1} shows that one may increase the number of test subjects twofold, using only $m=m_b$ tests. The increase is achieved by using a carefully chosen multiplier for the second block. More precisely, this choice of $\alpha_2$ satisfies two properties. First, since $\alpha_2\succ_{\boldsymbol{\eta}}\alpha_1$, none of the two columns of $\mathbf{C}$ have the same syndrome, and therefore can be uniquely distinguished. Second, the fact that $\alpha_1+\alpha_2\succ_{\boldsymbol{\eta}}\alpha_2\succ_{\boldsymbol{\eta}}\alpha_1$, ensures that if we can identify a column of $\mathbf{C}_b$ that corresponds to at least one defective, denoted by $\mathbf{x}_b$, it is possible to determine if $\{\alpha_1\mathbf{x}_b\}$, or $\{\alpha_2\mathbf{x}_b\}$, or $\{\alpha_1\mathbf{x}_b,\alpha_2\mathbf{x}_b\}$ are the columns of $\mathbf{C}$ that correspond to the defectives. These two properties, combined with the disjunctness property of $\mathbf{C}_b$, ensure that any collection of up to $d$ items has a unique syndrome after quantization, even in the presence of up to $e$ errors. This construction can be generalized to include concatenations of more than two matrices using the new families of quantized $B_h$ sequences and the SQLO$_s$ and SQLO$_l$ sequences, described next.

\section{SQ-separable codes using quantized $B_h$ sequences} \label{sec:bh}
We start by introducing quantized $B_h$ sequences which generalize the well known $B_h$ sequences from number theory. First, we define the standard $B_h$ sequences~\cite{HR83}.
\begin{defin}[\textbf{$B_h$ sequence}]
A finite sequence of positive integers $\mathcal{A}=\{\alpha_1,\alpha_2,\dots,\alpha_K\}$ is a $B_h$ sequence if 
$\forall \mathcal{A}_1,\mathcal{A}_2\subseteq\mathcal{A}$ such that $\mathcal{A}_1\neq\mathcal{A}_2$, $|\mathcal{A}_1|=|\mathcal{A}_2|= h$, one has $\sum_{\alpha_i\in\mathcal{A}_1}\alpha_i\neq\sum_{\alpha_i\in\mathcal{A}_2}\alpha_i$. 
\end{defin}

Similar to the classical $B_h$ sequences which require distinct subset sums of cardinality $h$, in quantized $B_h$ sequences we require that the \textit{quantized} sums of subsets of size \textit{up to} $h$ be distinct. These sequences can be used to generalize Thm.~\ref{const1} to construct SQ-separable codes.

\begin{defin}[\textbf{Quantized $B_h$ sequence}]\label{SQBh}
A finite sequence of positive integers $\mathcal{A}=\{\alpha_1,\alpha_2,\dots,\alpha_K\}$ is called a quantized $B_h$ sequence with respect to $\boldsymbol{\eta}$ if 
\begin{enumerate}
\item $\alpha_K\succ_{\boldsymbol{\eta}}\alpha_{K-1}\succ_{\boldsymbol{\eta}}\dots\succ_{\boldsymbol{\eta}}\alpha_1\succ_{\boldsymbol{\eta}}0$ (i.e., all elements of $\mathcal{A}$ lie in different quantization bins).
\item $\forall \mathcal{A}_1,\mathcal{A}_2\subseteq\mathcal{A}$ such that $\mathcal{A}_1\neq\mathcal{A}_2$, $|\mathcal{A}_1|\leq h$ and $|\mathcal{A}_2|\leq h$, one either has $\sum_{\alpha_i\in\mathcal{A}_1}\alpha_i\succ_{\boldsymbol{\eta}}\sum_{\alpha_i\in\mathcal{A}_2}\alpha_i$ or $\sum_{\alpha_i\in\mathcal{A}_2}\alpha_i\succ_{\boldsymbol{\eta}}\sum_{\alpha_i\in\mathcal{A}_1}\alpha_i$ (the sums of elements of distinct subsets lie in different quantization bins). 
\end{enumerate}
\end{defin}
Intuitively, we require that all the elements of the sequence are located in different quantization bins, none of them is in the same bin as $0$, and in addition, all the sums that are formed by adding elements of subsets of cardinality at most $h$ fall into different bins. 
Note that when $K=2$, setting $\alpha_1=\eta_1$ and $\alpha_2=\max\{\eta_2,\eta_3-\eta_1\}$ as was done in Thm.~\ref{const1} ensures that the condition in the aforementioned definition are met.

\begin{remark}\label{rem1}
Note that the cardinality of a finite quantized $B_h$ sequence may be smaller than the value of $h$. For example, $\mathcal{A}=\{\eta_1\}$ is a quantized $B_h$ sequence with respect to $\boldsymbol{\eta}$, for any $h\in\mathbb{Z}^+$. However, one seeks to find the densest such sequence given an upper bound on the values of its largest element.
\end{remark}

Quantized $B_h$ sequences can be used to construct SQ-separable codes as shown in the next theorem.

\begin{theorem}\label{const_Bh}
Fix a binary $d$-disjunct code matrix $\mathbf{C}_b$ of dimensions $m_b\times n_b$, capable of correcting up to $e$ errors. Let $\mathcal{A}=\{\alpha_1,\alpha_2,\dots,\alpha_K\}$ be a quantized $B_d$ sequence with respect to $\boldsymbol{\eta}$. Form a matrix $\mathbf{C}$ of length $m=m_b$ and size $n=Kn_b$ by concatenating $K$ matrices $\mathbf{C}_i=\alpha_i\mathbf{C}_b$, $1\leq i\leq K$, horizontally. The constructed code is a $[q;Q;\boldsymbol{\eta};(1\!:\!d);e]$-SQ-separable code with $q=\alpha_K+1$.\end{theorem} 

\begin{IEEEproof}
In order to show that the constructed code is $[q;Q;\boldsymbol{\eta};(1\!:\!d);e]$-SQ-separable, we consider two distinct sets of codewords $\mathcal{X}_1$ and $\mathcal{X}_2$ that satisfy $1\leq|\mathcal{X}_1|,|\mathcal{X}_2|\leq d$. The idea is to show that the syndrome of these two sets contain at least $2e+1$ different entries. Without loss of generality, we assume that $|\mathcal{X}_1|\leq|\mathcal{X}_2|$. Since the two sets are distinct, one must have $\mathcal{X}_2\backslash\mathcal{X}_1\neq\varnothing$, and therefore we choose $\mathbf{z}_r\in\mathcal{X}_2\backslash\mathcal{X}_1$. By construction, $\mathbf{z}_r=\alpha_r\mathbf{z}_b$ for some binary codeword $\mathbf{z}_b$ in $\mathbf{C}_b$ and some $\alpha_r\in\mathcal{A}$. 

For the fixed binary codeword $\mathbf{z}_b$, let $\mathcal{Z}$, be the set of codewords of $\mathbf{C}$ generated by multiplying $\mathbf{z}_b$ with the elements of $\mathcal{A}$. Let $\mathcal{Z}_{1}=\mathcal{X}_1\cap\mathcal{Z}$ and $\mathcal{Z}_{2}=\mathcal{X}_2\cap\mathcal{Z}$, be the set of codewords with the same support as $\mathbf{z}_b$ in $\mathcal{X}_1$ and $\mathcal{X}_2$, respectively. Also, let $\mathcal{A}_{\mathcal{Z}_1}\subset\mathcal{A}$ and $\mathcal{A}_{\mathcal{Z}_2}\subseteq\mathcal{A}$ be the set of coefficients used to form the codewords in $\mathcal{Z}_1$ and $\mathcal{Z}_2$, respectively.
Given that $\mathcal{A}$ is a quantized $B_d$ sequence, we have to separately consider two different scenarios.

\textbf{Case 1:} $\sum_{\alpha_i\in\mathcal{A}_{\mathcal{Z}_2}}\alpha_i\succ_{\boldsymbol{\eta}}\sum_{\alpha_i\in\mathcal{A}_{\mathcal{Z}_1}}\alpha_i$.

By construction of $\mathbf{C}$ and Def.~\ref{CGTdisjunct}, there exists a set of coordinates $\mathcal{R}_r$ of size at least $2e+1$, such that $\forall k\in\mathcal{R}_r$, 
\begin{align}
\begin{cases}
\mathbf{z}_r(k)=\alpha_r,\\\nonumber
\mathbf{x}(k)=0\hspace{30pt}\forall \mathbf{x}\in\mathcal{X}_1\backslash\mathcal{Z}_1.
\end{cases}
\end{align}
Consequently, $\forall k\in\mathcal{R}_r$ we have the following sequence of inequalities:
\begin{align}\label{th3_1}
\mathbf{y}_{\!_{\mathcal{X}_2}}(k)&\geq\mathbf{y}_{\!_{\mathcal{Z}_2}}(k)\\\label{th3_2}
&>\mathbf{y}_{\!_{\mathcal{Z}_1}}(k)\\\label{th3_3}
&=\mathbf{y}_{\!_{\mathcal{X}_1}}(k)
\end{align}
where \eqref{th3_1} follows since $\mathcal{Z}_2\subseteq\mathcal{X}_2$, \eqref{th3_2} follows since $\sum_{\alpha_i\in\mathcal{A}_{\mathcal{Z}_2}}\alpha_i\succ_{\boldsymbol{\eta}}\sum_{\alpha_i\in\mathcal{A}_{\mathcal{Z}_1}}\alpha_i$, and \eqref{th3_3} follows since  $\mathbf{x}(k)=0,\  \forall \mathbf{x}\in\mathcal{X}_1\backslash\mathcal{Z}_1$.

 \textbf{Case 2:} $\sum_{\alpha_i\in\mathcal{A}_{\mathcal{Z}_1}}\alpha_i\succ_{\boldsymbol{\eta}}\sum_{\alpha_i\in\mathcal{A}_{\mathcal{Z}_2}}\alpha_i$.
 
In this case, we cannot use the set of coordinates $\mathcal{R}_r$, since \eqref{th3_2} no longer holds. On the other hand, this case happens only if $\mathcal{A}_{\mathcal{Z}_1}\backslash\mathcal{A}_{\mathcal{Z}_2}\neq\varnothing$. Consequently, one has $\mathcal{Z}_1\backslash\mathcal{Z}_2\neq\varnothing$; let $\mathbf{z}_s\in\mathcal{Z}_1\backslash\mathcal{Z}_2$, where $\mathbf{z}_s=\alpha_s\mathbf{z}_b$ for some $\alpha_s\in\mathcal{A}_{\mathcal{Z}_1}$. Similar to case 1, by considering on $\mathcal{X}_2$ instead of $\mathcal{X}_1$, there exists a set of coordinates $\mathcal{R}_s$ of size at least $2e+1$, such that $\forall k\in\mathcal{R}_s$, 
\begin{align}
\begin{cases}
\mathbf{z}_s(k)=\alpha_s,\\\nonumber
\mathbf{x}(k)=0\hspace{30pt}\forall \mathbf{x}\in\mathcal{X}_2\backslash\mathcal{Z}_2.
\end{cases}
\end{align}
As a result, the following inequalities hold:
\begin{align}\label{th3_4}
\mathbf{y}_{\!_{\mathcal{X}_1}}(k)&\geq\mathbf{y}_{\!_{\mathcal{Z}_1}}(k)\\\label{th3_5}
&>\mathbf{y}_{\!_{\mathcal{Z}_2}}(k)\\\label{th3_6}
&=\mathbf{y}_{\!_{\mathcal{X}_2}}(k),
\end{align}
where \eqref{th3_4} follows since $\mathcal{Z}_1\subseteq\mathcal{X}_1$, \eqref{th3_5} follows since $\sum_{\alpha_i\in\mathcal{A}_{\mathcal{Z}_1}}\alpha_i\succ_{\boldsymbol{\eta}}\sum_{\alpha_i\in\mathcal{A}_{\mathcal{Z}_2}}\alpha_i$, and \eqref{th3_6} follows since  $\mathbf{x}(k)=0,\  \forall \mathbf{x}\in\mathcal{X}_2\backslash\mathcal{Z}_2$. Note that even though $|\mathcal{X}_1|\leq|\mathcal{X}_2|$, unlike for Case 1, we have $\mathbf{y}_{\!_{\mathcal{X}_1}}(k)>\mathbf{y}_{\!_{\mathcal{X}_2}}(k)$ for all $k\in\mathcal{R}_s$.

\end{IEEEproof}

\subsection{Fundamental limits and constructions of quantized $B_h$ sequences}

Quantized $B_h$ sequences ensure that a set of integers and their subset sums are placed into different quantization bins. As a result, for a fixed set of $Q$ thresholds $\boldsymbol{\eta}$, the existence of quantized $B_h$ sequences with a predetermined cardinality $K$ depends on the thresholds. As mentioned in Remark~\ref{rem1}, the cardinality of a quantized $B_h$ sequence may be smaller than $h$. For example, one can always choose $\mathcal{A}=\{\eta_1\}$ as a quantized $B_h$ sequence with $K=1$. For the case of $K=2$, the sequence $\mathcal{A}=\{\eta_1,\max\{\eta_2,\eta_3-\eta_1\}\}$ used in Thm.~\ref{const1} is a quantized $B_h$ sequence with respect to $\boldsymbol{\eta}$ as long as $Q\geq 4$ and $\eta_Q>\eta_1+\max\{\eta_2,\eta_3-\eta_1\}$. These two examples imply that for any set of thresholds, there always exists a quantized $B_h$ sequence, which in the worst case scenario has cardinality $K=1$. 

We discuss next constructions of quantized $B_h$ sequences with $K>2$. From a practical perspective, and given that in most applications $q$ cannot be too large, a greedy algorithm for finding a quantized $B_h$ sequence is the simplest constructive approach. In the greedy approach, one starts with $\alpha_1=\eta_1$; then, given the first $i$ elements of the sequence, to find $\alpha_{i+1}$, one increases the value of $\alpha_i$ until the properties of the quantized $B_h$ sequence are satisfied. 

Although this method works for small values of $K$, for large values of $K$ this procedure has a high computational complexity. Alternatively, one can use standard subset-sum distinct sequences\footnote{A subset-sum distinct sequence is a sequence of positive integers such that the sum of the elements of its subsets are distinct.}~\cite{HR83}, and generalizations of standard $B_h$ sequences to construct a family of quantized $B_h$ sequences as described in the next theorem.

\begin{theorem}\label{qbh_const}
Consider a SQGT model with thresholds $\boldsymbol{\eta}=[0,\eta_1,\eta_2,\dots,\eta_Q]^T$; $\forall s:1\leq s\leq Q$, and let $g_s=\max_{i:1\leq i\leq s}\eta_i-\eta_{i-1}$ be the largest gap of the first $s$ thresholds. Let $\mathcal{B}=\{\beta_1<\beta_2<\dots\}$ be a sequence for which all the subset sums of at most $h$ elements are distinct. For a fixed $s$, $2\leq s\leq Q$, let $K_s$ be the largest positive integer that satisfies $\eta_s> g_s\sum_{i=\max\{1,K_s-h\}}^{K_s} \beta_i$. Then all the sequences of the form $\mathcal{A}_s = \left\{g_s\:\beta_1, g_s\:\beta_2, \dots, g_s\:\beta_{K_s}\right\}$ are quantized $B_h$ sequences with respect to $\boldsymbol{\eta}$. 

\end{theorem}
\begin{IEEEproof}
First note that $\eta_s> g_s\sum_{i=\max\{1,K_s-h\}}^{K_s} \beta_i$ guarantees that the sum of up to $h$ members of $\mathcal{A}_s$ never exceeds the largest threshold $\eta_Q$. Now, fix a value of $s:1\leq s\leq Q$, and consider any two distinct sets $\mathcal{A}_1,\mathcal{A}_2\subseteq\mathcal{A}_s$, $|\mathcal{A}_1|\leq h$ and $\mathcal{A}_2\leq h$, which are obtained by multiplying the elements of $\mathcal{B}_1\subseteq\mathcal{B}$ and $\mathcal{B}_2\subseteq\mathcal{B}$ with $g_s$, respectively. Suppose $f_{\boldsymbol{\eta}}\left(\sum_{\alpha_i\in\mathcal{A}_1}\alpha_i\right)=f_{\boldsymbol{\eta}}\left(\sum_{\alpha_i\in\mathcal{A}_2}\alpha_i\right)$; as a result, there exists $r$, $1\leq r\leq s$, such that $\eta_{r-1}\leq \sum_{\alpha_i\in\mathcal{A}_1}\alpha_i<\eta_{r}$ and $\eta_{r-1}\leq \sum_{\alpha_i\in\mathcal{A}_2}\alpha_i<\eta_{r}$. 
Consequently,
\begin{align}\label{cont1}
\left|\sum_{\alpha_i\in\mathcal{A}_1}\alpha_i-\sum_{\alpha_i\in\mathcal{A}_2}\alpha_i\right|\leq\eta_{r}-\eta_{r-1}-1< g_s.
\end{align}
However, since all the sums of up to $h$ elements of $\mathcal{B}$ are distinct and $|\mathcal{B}_1|\leq h$ and $|\mathcal{B}_2|\leq h$, $\left|\sum_{\beta_i\in\mathcal{B}_1}\beta_i-\sum_{\beta_i\in\mathcal{B}_2}\beta_i\right|\geq 1$. Consequently,
\begin{align}
\left|\sum_{\alpha_i\in\mathcal{A}_1}\alpha_i-\sum_{\alpha_i\in\mathcal{A}_2}\alpha_i\right|=g_s\left|\sum_{\beta_i\in\mathcal{B}_1}\beta_i-\sum_{\beta_i\in\mathcal{B}_2}\beta_i\right|\geq g_s,
\end{align}
which contradicts~\eqref{cont1}. 
\end{IEEEproof}

Given this theorem, one can construct quantized $B_h$ sequences using the sequences mentioned in the theorem or the more strict subset-sum distinct sequences, for which many constructions are known in the literature~\cite{HR83,L88,B97}. One should note that for a fixed value of $K$, the construction of quantized $B_h$ sequences described in this theorem may not generate the densest sequence; however, this construction has the important property that it applies to any set of thresholds and only depends on a condition that can be easily verified given the thresholds.

\begin{remark} All the subset-sums consisting of at most $h$ elements of a quantized $B_h$ sequence must fall into different quantization bins; since there are $Q$ such bins, the following bounds on the number of elements of a quantized $B_h$ sequence hold: 
Let $\mathcal{A}$ be a finite quantized $B_h$ sequence with respect to $\boldsymbol{\eta}$ such that $|\mathcal{A}|= K$. If $K\leq h$, then $K\leq\log_2Q$. On the other hand, if $K> h$, then $\sum_{i=0}^h{K\choose i}\leq Q$.
\end{remark}

\begin{remark}\label{rem5}
Let $\mathcal{B}$ be a subset-sum distinct sequence (i.e. a sequence such that all its subsets sum up to distinct values). Assume that a positive integer $K$ satisfies the condition in Thm.~\ref{qbh_const}; then, this theorem can be used to construct a quantized $B_h$ sequence $\mathcal{A}$, $|\mathcal{A}|=K$, using $\mathcal{B}=\{\beta_1,\beta_2,\dots,\beta_K\}$. There exist a large body of literature describing constructive bounds on $\beta_K$~\cite{L88},~\cite{B97}. All bounds are of the form $\beta_K\leq c2^{K}$, where $c<1$ is a constant that depends on the construction (e.g. $c=0.22002$ in~\cite{B97}). Given a bound of this form, one has $\alpha_K<cg2^{K}$, where $g$ is the largest gap for the first $K$ thresholds. 
\end{remark} 

The aforementioned bound is exponential in $K$, where the base of the exponential equals $2$. In Lemma~\ref{prop_recur}, we will prove an upper bound on $\alpha_K$ in which the base of the exponential function is strictly smaller than $2$. Although this bound applies to SQLO$_s$ sequences, given that any SQLO$_s$ sequence is also a quantized $B_h$ sequence, it can be considered an upper bound for quantized $B_h$ sequences as well. This implies that the bound in Lemma~\ref{prop_recur} is asymptotically tighter compared to aforementioned bound.

\subsection{A decoding algorithm for SQGT codes constructed using quantized $B_h$ sequences}
We describe next a decoding algorithm for codes constructed using Theorem~\ref{const_Bh}. Let $\mathcal{D}$ denote the set of codewords of $\mathbf{C}$ corresponding to the defectives. Also, let $\mathcal{X}_{\mathcal{D}}$ be the set of binary codewords each corresponding to the support of at least one codeword in $\mathcal{D}$; clearly, $|\mathcal{X}_{\mathcal{D}}|\leq|\mathcal{D}|\leq d$. The following example illustrates the relationship between $\mathcal{D}$ and $\mathcal{X}_{\mathcal{D}}$.

\begin{example}
As an example, suppose that in a SQGT system $\mathcal{D}=\{[2,0,2,2]^T,[6,0,6,6]^T,[2,0,2,0]^T\}$; in this case one has $\mathcal{X}_{\mathcal{D}}=\{[1,0,1,1]^T,[1,0,1,0]^T\}$. 
\end{example}

The decoding procedure is performed in three steps. The idea is to use the disjunctness property of binary disjunct matrices and the property of quantized $B_h$ sequences to first recover the set $\mathcal{X}_{\mathcal{D}}$ in Step 1, and then use this set to recover $\mathcal{D}$ in Steps 2 and 3. The steps of the decoding algorithm are listed in Algorithm~1.

\begin{table}
\begin{center}
\line(1,0){469}
\end{center}
\vspace{-8pt}
\textbf{Algorithm 1: Dec-QBh} 
\vspace{-10pt}
\begin{center}
\line(1,0){469}
\end{center}
\textbf{Input:} $\mathbf{y}\in[Q]^{m}$, $\mathbf{C}_b\in[2]^{m\times \frac{n}{K}}$, $\boldsymbol{\eta}$, $\mathcal{A}$, $e\geq0$ \\
\textbf{Output:} $\hat{\mathcal{D}}$\\

\textbf{Step 1:} Initialize $\mathcal{X}\leftarrow\varnothing$ and $\hat{\mathcal{D}}\leftarrow\varnothing$\\
\hspace*{20pt} \textbf{For} $i=1,2,\dots,\frac{n}{K}$ \textbf{do}\\
\hspace*{40pt} \textbf{If} the number of coordinates $j$ for which the $i$-th codeword of $\mathbf{C}_b$ does not satisfy \hspace*{60pt}$\mathbf{x}_i(j)\leq\mathbf{y}(j)$ is at most equal to $e$, set $\mathcal{X}\leftarrow\mathcal{X}\cup\{\mathbf{x}_i\}$.\\
\hspace*{40pt} \textbf{End}\\
\hspace*{20pt} \textbf{End}\\

\textbf{Step 2:}\\
\hspace*{20pt} Form $\mathcal{B}$ the ordered list of the distinct sums of elements of subsets of $\mathcal{A}$ with cardinality at most $d$ and their corresponding subsets.\\

\textbf{Step 3:} \\
\hspace*{20pt} Form $\mathbf{u}_{\!_{\mathcal{D}}}$ such that $\mathbf{u}_{\!_{\mathcal{D}}}(j)$ is the upper threshold of the quantization bin in which $\mathbf{y}(j)$ lies.\\
\hspace*{20pt} \textbf{For} $i=1,2,\dots,|\mathcal{X}|$ \textbf{do}\\
\hspace*{40pt} Find $\beta_l$, the largest element of $\mathcal{B}$ such that the number of coordinates $j$ for which \\\hspace*{40pt} $\beta_l\mathbf{x}_i(j)<\mathbf{u}_{\!_\mathcal{D}}(j)$ is not satisfied is at most $e$.\\
\hspace*{40pt} Let $\mathcal{A}_{i,l}\subseteq\mathcal{A}$ be the set with the sum equal to $\beta_l$. \\
\hspace*{40pt} {Set $\hat{\mathcal{D}}_i\leftarrow\{\textnormal{codewords of }\mathbf{C} \textnormal{ of the form } \mathbf{z}=\alpha\mathbf{x}_i,\  \forall \alpha\in\mathcal{A}_{i,l}       \}$}.\\
\hspace*{20pt} \textbf{End}\\

\textbf{Return} $\hat{\mathcal{D}}=\bigcup_i\hat{\mathcal{D}}_i$
\vspace{-8pt}
\begin{center}
\line(1,0){469}
\end{center}
\vspace{-20pt}
\end{table}

\begin{theorem}
Algorithm Dec-QBh is capable of identifying up to $d$ defectives in the presence of at most $e$ errors in the vector of test results $\mathbf{y}$.
\end{theorem}
\begin{IEEEproof}
In the first step of the algorithm, and for each codeword of the binary codebook $\mathbf{C}_b$, we count the number of coordinates for which the test result is smaller than the corresponding entry of the codeword. In order to show that the set $\mathcal{X}$ recovered in Step 1 is equal to $\mathcal{X}_{\mathcal{D}}$, we first show that $\mathcal{X}\supseteq\mathcal{X}_\mathcal{D}$. Each codeword in $\mathcal{D}$ can be written as $\mathbf{z}_i=\alpha\mathbf{x}_i$, $1\leq i\leq|\mathcal{D}|$, for some $\alpha\in\mathcal{A}$ and some binary codeword $\mathbf{x}_i$ in $\mathcal{X}_\mathcal{D}$. We need to show that if $\mathbf{x}_i\in\mathcal{X}_{\mathcal{D}}$, then $\mathbf{x}_i\in\mathcal{X}$, or equivalently, the number of coordinates $j$ for which
\begin{align}\label{decoding_eq1}
\mathbf{x}_i(j)\leq\mathbf{y}(j)
\end{align} 
is not satisfied is at most $e$.
All the entries of $\mathbf{y}$ which are not erroneous are equal to the corresponding entries of the syndrome of defectives $\mathbf{y}_{\!_\mathcal{D}}$. As a result,~\eqref{decoding_eq1} is trivially satisfied for entries of $\mathbf{x}_i$ that are equal to zero, since for these entries $\mathbf{y}_{\!_\mathcal{D}}$ is equal to zero and an error can only increase the corresponding coordinate in $\mathbf{y}$. On the other hand, since $\mathcal{A}$ is a quantized $B_d$ sequence, its smallest element satisfies $\alpha_1\geq\eta_1$. Consequently, a nonzero entry of $\mathbf{x}_i$ results in a nonzero entry in $\mathbf{y}_{\!_\mathcal{D}}$, which is a nonzero entry in $\mathbf{y}$ unless an error occurs; since the nonzero entries of $\mathbf{x}_i$ are equal to $1$ (the smallest positive integer) and there are at most $e$ errors, condition~\eqref{decoding_eq1} is satisfied for all except up to $e$ nonzero entries. Consequently, $\mathcal{X}\supseteq\mathcal{X}_\mathcal{D}$.

Next, we show that if $\mathbf{x}_i\in\mathcal{X}$, then $\mathbf{x}_i\in\mathcal{X}_{\mathcal{D}}$, or equivalently $\mathcal{X}\subseteq\mathcal{X}_{\mathcal{D}}$. Suppose this is not true and let $\mathbf{x}\in\mathcal{X}\backslash\mathcal{X}_{\mathcal{D}}$. Since $\mathbf{C}_b$ is a binary disjunct matrix and $|\mathcal{X}_{\mathcal{D}}|\leq d$, then there exists a set of coordinates $\mathcal{R}$ such that $|\mathcal{R}|\geq 2e+1$ and $\forall j\in\mathcal{R}$ one has $\mathbf{x}(j)=1$ while $\mathbf{x}_i(j)=0$, $\forall \mathbf{x}_i\in\mathcal{X}_{\mathcal{D}}$. Consequently, $\forall j\in\mathcal{R}$, one has $\mathbf{y}_{\!_\mathcal{D}}(j)=0$, which implies that $\mathbf{y}(j)=0$ unless an error occurred. Since there are at most $e$ errors, $\mathbf{x}(j)>\mathbf{y}(j)$ for at least $e+1$ coordinates, which implies that $\mathbf{x}\nin\mathcal{X}$. This contradicts the starting assumption. Hence, $\mathcal{X}\subseteq\mathcal{X}_{\mathcal{D}}$.

Now given that Step 1 recovered the set $\mathcal{X}=\mathcal{X}_{\mathcal{D}}$, we only need to show that Step 3 recovers $\mathcal{D}$ given $\mathcal{X}_{\mathcal{D}}$. For each $\mathbf{x}_i\in\mathcal{X}_{\mathcal{D}}$, let $\mathcal{A}_{i,t}$ be the ``true'' set of coefficients used to generate the codewords in $\mathcal{D}$ with the same support as $\mathbf{x}_i$. Also, let $\beta_t=\sum_{\alpha\in\mathcal{A}_{i,t}}\alpha$ be the sum of these coefficients. Since the error-free entries of $\mathbf{y}$ are equal to $\mathbf{y}_{\!_\mathcal{D}}$,  then for all $1\leq j\leq m$, one has $\beta_t\mathbf{x}_i(j)<\mathbf{u}_{\mathcal{D}}(j)$ unless an error occurred in the $j$-th coordinate. Since there are at most $e$ errors, there are at most $e$ coordinates for which this condition is not satisfied. As a result, $\beta_l\geq\beta_t$. 

In order to complete the proof, we show that no value of $\beta'\in\mathcal{B}$ such that $\beta'>\beta_t$ satisfies the condition in Step 3 and hence conclude that $\beta_l\leq\beta_t$. From the disjunctness property of $\mathbf{C}_b$, there exists a set of coordinates $\mathcal{R}_i$ such that $|\mathcal{R}_i|\geq2e+1$ and $\forall j\in\mathcal{R}_i$, $\mathbf{x}_i(j)=1$, while all other codewords in $\mathcal{X}_{\mathcal{D}}$ have the value zero at that coordinate. As a result, $\forall j\in\mathcal{R}_i$,
\begin{align}\nonumber
\sum_{\mathbf{z}\in\mathcal{D}}\mathbf{z}(j)=\sum_{\alpha\in\mathcal{A}_{i,t}}\alpha\mathbf{x}_i(j)=\sum_{\alpha\in\mathcal{A}_{i,t}}\alpha=\beta_t.
\end{align}
Since there are at most $e$ errors in $\mathbf{y}$, there exists a set of coordinates $\mathcal{R}'_i\subseteq\mathcal{R}_i$ with $|\mathcal{R}'_i|\geq e+1$, such that $\forall j\in\mathcal{R}'_i$,
\begin{align}\nonumber
\mathbf{u}_{\mathcal{D}}(j)>\sum_{\mathbf{z}\in\mathcal{D}}\mathbf{z}(j)=\beta_t.
\end{align}
Consider $\beta'\in\mathcal{B}$ such that $\beta'>\beta_t$. Since $\mathcal{A}$ is a quantized $B_d$ sequence, $\beta'\succ_{\boldsymbol{\eta}}\beta_t$ implies that $\forall j\in\mathcal{R}'_i$ one has $\beta'\geq\mathbf{u}_{\mathcal{D}}(j)>\beta_t$. Given $|\mathcal{R}'_i|\geq e+1$, the condition in Step 3 is not satisfied for such a choice of $\beta'$ and hence $\beta_t\geq\beta_l$. As a result, Step 3 uniquely recovers $\beta_l=\beta_t$ which corresponds to the set $\mathcal{A}_{i,t}$. Consequently, $\hat{\mathcal{D}}=\mathcal{D}$ in the presence of up to $e$ errors in the vector of test results, as claimed. 
\end{IEEEproof}

\begin{remark}
The computational complexity of Algorithm~1 is equal to $O(\frac{mn}{K}+2^K(K+md))$. The computational complexity of Step 1 is $O(\frac{mn}{K})$. The second step requires $\sum_{i=1}^{\min\{d,K\}}{K\choose i}(i-1)=O(K2^K)$ summations. Finally, the computational complexity of Step 3 is $O(dm2^K)$.
\end{remark}

Due to the exponential growth of the the computational complexity of the decoding algorithm with $K$, the codes constructed using quantized $B_h$ sequences are most suitable for small values of $K$. On the other hand, for larger values of $K$, we introduce two other families of sequences that lead to codes with significantly smaller decoding complexity.

\section{SQ-separable codes using SQLO$_s$ sequences} \label{sec:sqlos}
As discussed earlier, the codes constructed using quantized $B_h$ sequences have a decoding algorithm with computational complexity $O(\frac{mn}{K}+2^K(K+md))$. Although for small values of $K$ the dominant term is $\frac{mn}{K}$, for large values of $K$ the exponential growth of the complexity with $K$ is problematic. In this section we introduce the notion of SQLO$_s$ sequences and use them to construct SQ-separable codes with a decoding algorithm that has computational complexity linear in $K$.

\begin{defin}[\textbf{SQLO$_s(\boldsymbol{\eta},h)$ sequences}]\label{SQLO}
Given a set of thresholds $\boldsymbol{\eta}$, a sequence of positive integers $\mathcal{A}=\{\alpha_1,\alpha_2,\dots,\alpha_K\}$ is termed a SQLO$_s(\boldsymbol{\eta},h)$ sequence if
\begin{enumerate}
\item $\alpha_K\succ_{\boldsymbol{\eta}}\alpha_{K-1}\succ_{\boldsymbol{\eta}}\dots\alpha_2\succ_{\boldsymbol{\eta}}\alpha_1\succ_{\boldsymbol{\eta}}0$ (i.e., all elements of $\mathcal{A}$ lie in different quantization bins).
\item For any two distinct nonempty nested subsets $\mathcal{A}_1\subset\mathcal{A}_2\subseteq\mathcal{A}$ such that $|\mathcal{A}_1|\leq h$ and $|\mathcal{A}_2|\leq h$, one has $\sum_{\alpha_i\in\mathcal{A}_2}\alpha_i\succ_{\boldsymbol{\eta}}\sum_{\alpha_i\in\mathcal{A}_1}\alpha_i$ (i.e., the sums of elements of nested subsets fall into different quantization bins).
\item For any two distinct nonempty subsets $\mathcal{A}_1,\mathcal{A}_2\subseteq\mathcal{A}$ that are not nested and $|\mathcal{A}_1|\leq h$ and $|\mathcal{A}_2|\leq h$, one has $\sum_{\alpha_i\in\mathcal{A}_2}\alpha_i\succ_{\boldsymbol{\eta}}\sum_{\alpha_i\in\mathcal{A}_1}\alpha_i$ whenever $\exists \alpha\in\mathcal{A}_2\backslash\mathcal{A}_1$ such that $\alpha\succ_{\boldsymbol{\eta}} \alpha'$, $\forall \alpha'\in\mathcal{A}_1\backslash\mathcal{A}_2$ (i.e., two subsets that are not nested are ordered based on their largest distinct element).
\end{enumerate}
\end{defin}
The properties above induce a partial order on the subsets of a SQLO$_s$ sequence.

The SQLO$_s$ properties for $K=2$ and $h\geq 2$ simply translates to $\alpha_2+\alpha_1\succ_{\boldsymbol{\eta}}\alpha_2\succ_{\boldsymbol{\eta}}\alpha_1\succ_{\boldsymbol{\eta}}0$, while for $K=3$ and $h\geq 3$ it translates into $\alpha_3+\alpha_2+\alpha_1\succ_{\boldsymbol{\eta}}\alpha_3+\alpha_2\succ_{\boldsymbol{\eta}}\alpha_3+\alpha_1\succ_{\boldsymbol{\eta}}\alpha_3\succ_{\boldsymbol{\eta}}\alpha_2+\alpha_1\succ_{\boldsymbol{\eta}}\alpha_2\succ_{\boldsymbol{\eta}}\alpha_1\succ_{\boldsymbol{\eta}}0$. The following example describes a SQLO$_s$ sequence.

\begin{example}
It can be easily verified that $\mathcal{A}=\{3,6,12\}$ is a SQLO$_s$ sequence with respect to the thresholds $\boldsymbol{\eta}=[0,3,6,\dots,24]^T$, since $f_{\boldsymbol{\eta}}(12+6+3)=7>f_{\boldsymbol{\eta}}(12+6)=6>f_{\boldsymbol{\eta}}(12+3)=5>f_{\boldsymbol{\eta}}(12)=4>f_{\boldsymbol{\eta}}(6+3)=3>f_{\boldsymbol{\eta}}(6)=2>f_{\boldsymbol{\eta}}(3)=1>0$.
\end{example}

SQLO$_s$ sequences obey a more stringent set of constraints compared to the quantized $B_h$ sequences. As a result, one is able to use these constraints to reduce the computational complexity of the decoder. In the next proposition, we show that in fact any SQLO$_s$ sequence is also a quantized $B_h$ sequence, but the converse is not necessarily true. 

\begin{prop}\label{prop_1}
A sequence of $K$ positive integers $\mathcal{A}$ is a SQLO$_s(\boldsymbol{\eta},h)$ sequence if and only if both of the following properties are satisfied:
\begin{enumerate}
\item $\mathcal{A}$ is a quantized $B_h$ sequence.
\item $\forall i:1\leq i\leq K$ and $\forall \mathcal{A}'\subseteq\{\alpha_1,\alpha_2,\dots,\alpha_{i-1}\}$ such that $|\mathcal{A}'|\leq h$, one has $\alpha_i\succ_{\boldsymbol{\eta}}\sum_{\alpha_j\in\mathcal{A}'}\alpha_j$.
\end{enumerate}

\end{prop}
\begin{IEEEproof}
First, we show that if $\mathcal{A}$ is a SQLO$_s(\boldsymbol{\eta},h)$ sequence, it satisfies properties 1 and 2. It is easy to see that since $\mathcal{A}$ is a SQLO$_s(\boldsymbol{\eta},h)$ sequence, it satisfies the first property of quantized $B_h$ sequences, i.e. $\alpha_K\succ_{\boldsymbol{\eta}}\alpha_{K-1}\succ_{\boldsymbol{\eta}}\dots\succ_{\boldsymbol{\eta}}\alpha_1\succ_{\boldsymbol{\eta}}0$.

 Let $\mathcal{A}_1$ and $\mathcal{A}_2$ be two arbitrary nonempty distinct subsets of $\mathcal{A}$ such that $|\mathcal{A}_1|\leq h$ and $|\mathcal{A}_2|\leq h$. We need to show that $\sum_{\alpha_i\in\mathcal{A}_1}\alpha_i\succ_{\boldsymbol{\eta}}\sum_{\alpha_i\in\mathcal{A}_2}\alpha_i$ or $\sum_{\alpha_i\in\mathcal{A}_2}\alpha_i\succ_{\boldsymbol{\eta}}\sum_{\alpha_i\in\mathcal{A}_1}\alpha_i$. If these two subsets are nested, i.e. if $\mathcal{A}_1\subset\mathcal{A}_2$ or $\mathcal{A}_2\subset\mathcal{A}_1$, from the second property of a SQLO$_s$ sequence, it follows that $\sum_{\alpha_i\in\mathcal{A}_2}\alpha_i\succ_{\boldsymbol{\eta}}\sum_{\alpha_i\in\mathcal{A}_1}\alpha_i$ or $\sum_{\alpha_i\in\mathcal{A}_1}\alpha_i\succ_{\boldsymbol{\eta}}\sum_{\alpha_i\in\mathcal{A}_2}\alpha_i$, respectively. Otherwise, the third property of a SQLO$_s(\boldsymbol{\eta},h)$ sequence ensures that $\mathcal{A}$ is a quantized $B_h$ sequence. On the other hand, from the third property of a SQLO$_s(\boldsymbol{\eta},h)$ sequence, one can directly conclude that the second property of the proposition holds.

Now we show that if $\mathcal{A}$ satisfies the two properties stated in the proposition, then it is a SQLO$_s(\boldsymbol{\eta},h)$ sequence. Since $\mathcal{A}$ is a quantized $B_h$ sequence, the first property of a SQLO$_s(\boldsymbol{\eta},h)$ sequence is automatically satisfied. 

Next, consider two distinct nonempty nested subsets $\mathcal{A}_1\subset\mathcal{A}_2\subseteq\mathcal{A}$ such that $|\mathcal{A}_1|\leq h$ and $|\mathcal{A}_2|\leq h$. Since $\sum_{\alpha_i\in\mathcal{A}_2}\alpha_i$ and $\sum_{\alpha_i\in\mathcal{A}_1}\alpha_i$ fall into different quantization bins, due to the second property of a quantized $B_h$ sequence, and since $\sum_{\alpha_i\in\mathcal{A}_2}\alpha_i>\sum_{\alpha_i\in\mathcal{A}_1}\alpha_i$, one has $\sum_{\alpha_i\in\mathcal{A}_2}\alpha_i\succ_{\boldsymbol{\eta}}\sum_{\alpha_i\in\mathcal{A}_1}\alpha_i$. Now consider two distinct nonempty subsets $\mathcal{A}_1,\mathcal{A}_2\subseteq\mathcal{A}$ that are not nested, such that $|\mathcal{A}_1|\leq h$ and $|\mathcal{A}_2|\leq h$. Assume that $\exists \alpha\in\mathcal{A}_2\backslash\mathcal{A}_1$ such that $\alpha\succ_{\boldsymbol{\eta}} \alpha'$, $\forall \alpha'\in\mathcal{A}_1\backslash\mathcal{A}_2$. In this case, it holds that
\begin{align}\nonumber
\sum_{\alpha_i\in\mathcal{A}_2}\alpha_i&\succ_{\boldsymbol{\eta}}\alpha\succ_{\boldsymbol{\eta}}\sum_{\alpha_i\in\mathcal{A}_1}\alpha_i,
\end{align}
where the last inequality follows from the second property of the proposition. This completes the proof of the proposition.
\end{IEEEproof}

As a result of the first condition in Proposition \ref{prop_1}, one can directly use a SQLO$_s(\boldsymbol{\eta},h)$ sequence instead of a quantized $B_h$ sequence to construct SQ-separable codes, as formally stated in the next theorem. In addition, the second property in Proposition \ref{prop_1} allows us to reduce the computational complexity of the decoder significantly. This is a consequence of the fact that superincreasing sequences\footnote{A superincreasing sequence is a sequence of positive integers such that each element of the sequence is at least as large as the sum of all the elements preceding it.} are knapsack-solvable in linear time~\cite{PHS03}. In other words, given an integer and a finite superincreasing sequence, it is possible to determine in linear time whether the integer can be expressed as a sum of distinct elements of the sequence, and if so to identify these elements~\cite{PHS03}.

\begin{theorem}\label{const_SQLO}
Fix a binary $d$-disjunct code matrix $\mathbf{C}_b$ of dimensions $m_b\times n_b$, capable of correcting up to $e$ errors. Let $\mathcal{A}=\{\alpha_1,\alpha_2,\dots,\alpha_K\}$ be a SQLO$_s(\boldsymbol{\eta},d)$ sequence. Form a matrix $\mathbf{C}$ of length $m=m_b$ and size $n=Kn_b$ by concatenating $K$ matrices $\mathbf{C}_i=\alpha_i\mathbf{C}_b$, $1\leq i\leq K$, horizontally. The constructed code is a $[q;Q;\boldsymbol{\eta};(1\!:\!d);e]$-SQ-separable code with $q=\alpha_K+1$.\end{theorem} 

\begin{IEEEproof}
The proof directly follows from Proposition~\ref{prop_1} and Thm.~\ref{const_Bh}. Since any SQLO$_s(\boldsymbol{\eta},d)$ sequence is a quantized $B_d$ sequence, Thm.~\ref{const_Bh} implies that the code $\mathbf{C}$ is a $[q;Q;\boldsymbol{\eta};(1\!:\!d);e]$-SQ-separable code with $q=\alpha_K+1$.
\end{IEEEproof}

\subsection{Fundamental limits and constructions of SQLO$_s$ sequences}

We discuss next construction methods and fundamental density limits for SQLO$_s$ sequences. 
Given a set of thresholds, a simple greedy algorithm can be used to find a SQLO$_s$ sequence by checking the properties in Def.~\ref{SQLO}, as demonstrated in the following example. 

\begin{example}
Suppose $\boldsymbol{\eta}=[0,2,5,6,10,13,15,16,18,21]^T$ and $h\geq K=3$; the greedy algorithm mentioned above produces $\mathcal{A}=\{2,5,11\}$. 
\end{example}

Alternatively, one can use the following theorem to construct SQLO$_s$ sequences using  superincreasing sequences. 

\begin{defin}
A sequence of positive integers $\mathcal{B}=\{\beta_1,\beta_2,\dots\}$ is called $h$-superincreasing if $\forall j>1$, $\beta_j>\sum_{i=\max\{1,j-h\}}^{j-1}\beta_i$.
\end{defin}

\begin{theorem}\label{SQLO1_const}
Consider a SQGT system with thresholds $\boldsymbol{\eta}=[0,\eta_1,\eta_2,\dots,\eta_Q]^T$; $\forall s:1\leq s\leq Q$, let $g_s=\max_{i:1\leq i\leq s}\eta_i-\eta_{i-1}$ be the largest gap of the first $s$ thresholds. Let $\mathcal{B}=\{\beta_1<\beta_2<\dots\}$ be a $h$-superincreasing sequence. For a fixed $s$, $2\leq s\leq Q$, let $K_s$ be the largest positive integer that satisfies $\eta_s> g_s\sum_{i=\max\{1,K_s-h\}}^{K_s} \beta_i$. Then all the sequences of the form $\mathcal{A}_s = \left\{g_s\:\beta_1, g_s\:\beta_2, \dots, g_s\:\beta_{K_s}\right\}$ are SQLO$_s(\boldsymbol{\eta},h)$ sequences. 

\end{theorem}
\begin{IEEEproof}
Let $\mathcal{A}_s$ be a fixed sequence satisfying the conditions of the theorem. First note that $\eta_s> g_s\sum_{i=\max\{1,K_s-h\}}^{K_s} \beta_i$ guarantees that the sum of up to $h$ members of $\mathcal{A}_s$ never exceeds the largest threshold $\eta_Q$. Next, we show that $\mathcal{A}_s$ is a quantized $B_h$ sequence. 
Fix a value of $s:1\leq s\leq Q$. Consider any two distinct sets $\mathcal{A}_1,\mathcal{A}_2\subseteq\mathcal{A}_s$, $|\mathcal{A}_1|\leq h$ and $|\mathcal{A}_2|\leq h$, which are obtained by multiplying the elements of $\mathcal{B}_1\subseteq\mathcal{B}$ and $\mathcal{B}_2\subseteq\mathcal{B}$ with $g_s$, respectively. Suppose that $f_{\boldsymbol{\eta}}\left(\sum_{\alpha_i\in\mathcal{A}_1}\alpha_i\right)=f_{\boldsymbol{\eta}}\left(\sum_{\alpha_i\in\mathcal{A}_2}\alpha_i\right)$; as a result, there exists an integer $r$, $1\leq r\leq s$, such that $\eta_{r-1}\leq \sum_{\alpha_i\in\mathcal{A}_1}\alpha_i<\eta_{r}$ and $\eta_{r-1}\leq \sum_{\alpha_i\in\mathcal{A}_2}\alpha_i<\eta_{r}$. 
Consequently,
\begin{align}\label{cont1_2}
\left|\sum_{\alpha_i\in\mathcal{A}_1}\alpha_i-\sum_{\alpha_i\in\mathcal{A}_2}\alpha_i\right|\leq\eta_{r}-\eta_{r-1}-1< g_s.
\end{align}
Since $\mathcal{B}_1\neq\mathcal{B}_2$, the set $(\mathcal{B}_1\cup\mathcal{B}_2)\backslash(\mathcal{B}_1\cap\mathcal{B}_2)$ is nonempty. Let $\beta_l$ be the largest element of this set, and without loss of generality assume that $\beta_l\in\mathcal{B}_1$.
Since $\mathcal{B}$ is a $h$-superincreasing sequence and $|\mathcal{B}_1|\leq h$ and $|\mathcal{B}_2|\leq h$, one has $\beta_l>\sum_{\beta_i\in\mathcal{B}_2}\beta_i$. This implies that $\sum_{\beta_i\in\mathcal{B}_1}\beta_i>\sum_{\beta_i\in\mathcal{B}_2}\beta_i$, or equivalently, that $\left|\sum_{\beta_i\in\mathcal{B}_1}\beta_i-\sum_{\beta_i\in\mathcal{B}_2}\beta_i\right|\geq 1$. Consequently,
\begin{align}
\left|\sum_{\alpha_i\in\mathcal{A}_1}\alpha_i-\sum_{\alpha_i\in\mathcal{A}_2}\alpha_i\right|=g_s\left|\sum_{\beta_i\in\mathcal{B}_1}\beta_i-\sum_{\beta_i\in\mathcal{B}_2}\beta_i\right|\geq g_s,
\end{align}
which contradicts~\eqref{cont1_2}; hence $\mathcal{A}_s$ is a quantized $B_h$ sequence. 

In order to complete the proof, we need to show that $\forall \alpha_j\in\mathcal{A}_s$, $1\leq j\leq K_s$, and $\forall \mathcal{A}_1\subseteq\{\alpha_1,\alpha_2,\dots,\alpha_{j-1}\}\subseteq\mathcal{A}_s$ such that $|\mathcal{A}_1|\leq h$, one has $\alpha_j\succ_{\boldsymbol{\eta}}\sum_{\alpha_i\in\mathcal{A}_1}\alpha_i$. Suppose this were not true and that $f_{\boldsymbol{\eta}}\left(\alpha_j\right)\leq f_{\boldsymbol{\eta}}\left(\sum_{\alpha_i\in\mathcal{A}_1}\alpha_i\right)=r$, $1\leq r\leq s$. As a result,
\begin{align}\label{SQLO1_proof1}
\alpha_j-\sum_{\alpha_i\in\mathcal{A}_1}\alpha_i<g_s.
\end{align}
Let $\beta_j=\frac{\alpha_j}{g_s}$ and let $\mathcal{B}_1\subseteq\mathcal{B}$ be the set which was used to construct $\mathcal{A}_1$. Since $\mathcal{B}$ is a $h$-superincreasing sequence and $|\mathcal{B}_1|\leq h$, one has
\begin{align}\label{SQLO1_proof2}
\beta_j>\sum_{\beta_i\in\mathcal{B}_1}\beta_i&\Rightarrow \beta_j-\sum_{\beta_i\in\mathcal{B}_1}\beta_i\geq 1.
\end{align}
By multiplying both sides of \eqref{SQLO1_proof2} by $g_s$, one has $\alpha_j-\sum_{\alpha_i\in\mathcal{A}_1}\alpha_i\geq g_s$ which contradicts~\eqref{SQLO1_proof1}. As a result, $\mathcal{A}_s$ is a SQLO$_s(\boldsymbol{\eta},h)$ sequence.
\end{IEEEproof}

Given this result, one can construct SQLO$_s(\boldsymbol{\eta},h)$ sequences using $h$-superincreasing sequences, as demonstrated in the following example. 

\begin{example}
For example, the sequence $\mathcal{B}=\{1,2,2^2,2^3,\dots\}$ is a superincreasing sequence, hence an $h$-superincreasing sequence for any value of $h$, and can be used to construct SQLO$_s(\boldsymbol{\eta},h)$ sequences. Given this sequence, one obtains a SQLO$_s(\boldsymbol{\eta},h)$ sequence such that $\alpha_K=g_s2^{K-1}$. 
\end{example}

Nevertheless, a simple construction based on recursive equations results in a better upper bound on the smallest value for $\alpha_K$, as described in Lemma~\ref{prop_recur}. 
We next state a theorem by Ostrovsky~\cite[Thm. 1.1.4]{P10} which we will use in the proof of Lemma~\ref{prop_recur}. The proof of this result can be found in~\cite[P. 3]{P10}.

\begin{lemma}\label{ost}
Let $P(x)=x^n-a_1x^{n-1}-\dots-a_n$, where all the coefficients $a_i$, $1\leq i\leq n$, are non-negative, and at least one is nonzero. If the greatest common divisor of the indices of the positive coefficients equals $1$, then the polynomial $P(x)$ has a unique positive root $r$; in addition, for any other root of this polynomial denoted by $r'$, one has $|r'|<r$.
\end{lemma}
Given this lemma, we will prove the following result.

\begin{lemma}\label{prop_recur}
Let $\gamma$ be the largest positive real root of the polynomial $g(x)=x^{h+1}-2x^h+1$. Also, assume that a positive integer $K$ satisfies the condition in Thm.~\ref{SQLO1_const}; then one can construct a SQLO$_s(\boldsymbol{\eta},h)$ sequence such that $\alpha_K=O_g\left(\gamma^{K}\right)$, where $g$ is the largest gap for the first $K$ thresholds, and $\gamma<2$. 
\end{lemma} 
\begin{IEEEproof}
We construct a sequence $\mathcal{B}$ as follows. First, $\forall 1\leq i\leq h$, we set $\beta_i=2^{i-1}$. Then, for $i>h$, we let $\beta_i=\beta_{i-1}+\beta_{i-2}+\dots+\beta_{i-h}+1$. Clearly, this sequence is a $h$-superincreasing sequence. The characteristic equation\footnote{The characteristic equation of a linear recurrence relation $a_i=c_1a_{i-1}+c_2a_{i-2}+\dots+c_ha_{i-h}$ is an equation of the form $x^h-c_1x^{h-1}-c_2x^{h-2}-\dots-c_h=0$, the roots of which are used to solve the recurrence. }  of this recurrence is of the form $f(x)=x^{h}-x^{h-1}-\dots-x-1=0$, which satisfies the condition of Lemma~\ref{ost}. In addition, the greatest common divisor of the indices of the positive coefficients is $1$, since all these coefficients are equal to $1$. Consequently, Lemma~\ref{ost} implies that this equation has a unique real positive root, $\gamma$, and that the absolute values of all the other roots are strictly smaller than $\gamma$. Consequently, $\beta_K=O\left(\gamma^{K}\right)$. Simplifying this equation by multiplying both sides by $(x-1)$, the equation becomes $x^{h+1}-2x^h+1=0$. Consequently, one has $\alpha_K=O_g\left(\gamma^{K}\right)$, where $\gamma$ is the largest positive real root of $g(x)$.

Next, we show that $\gamma<2$. Evaluating $g(x)=x^{h+1}-2x^h+1$ on the real axis reveals that this function has two local optima at $x=0$ and $x=\frac{2h}{h+1}$, and is monotonically increasing for $x>\frac{2h}{h+1}$. On the other hand, $g(2)>0$; in addition, for all $h\geq 1$, one has $2>\frac{2h}{h+1}$; consequently, $\forall x> 2$, $f(x)>f(2)>0$. As a result, the largest positive real solution to $g(x)=0$ is strictly smaller than $2$ for any finite value of $h$, i.e. $\gamma<2$.

\end{IEEEproof}

\subsection{A decoding algorithm for SQGT codes constructed using SQLO$_s$ sequences}

We next describe the Dec-SQLO$_s$ algorithm, the decoding procedure for codes based on SQLO$_s$ sequences. This algorithm comprises of two steps. The first step is identical to the first step of Algorithm 1. However, Steps 2 and 3 in Algorithm~1 are replaced by a single step which has a significantly lower computational complexity than steps 2 and 3. The steps of Dec-SQLO$_s$ are listed in Algorithm~2. The first step identifies the set $\mathcal{X}_{\mathcal{D}}$. Given this set, Step 2 identifies the set of defectives $\mathcal{D}$. In order to show that the second step can identify up to $d$ defectives in the presence of up to $e$ errors, we state the following lemma and proposition which we find useful for our subsequent proofs.

\begin{table}
\begin{center}
\line(1,0){469}
\end{center}
\vspace{-8pt}
\textbf{Algorithm 2: Dec-SQLO$_s$} 
\vspace{-10pt}
\begin{center}
\line(1,0){469}
\end{center}
\textbf{Input:} $\mathbf{y}\in[Q]^{m}$, $\mathbf{C}_b\in[2]^{m\times \frac{n}{K}}$, $\boldsymbol{\eta}$, $\mathcal{A}$, $e\geq0$ \\
\textbf{Output:} $\hat{\mathcal{D}}$\\

\textbf{Step 1:} Initialize $\mathcal{X}\leftarrow\varnothing$ and $\hat{\mathcal{D}}\leftarrow\varnothing$\\
\hspace*{0pt} \textbf{For} $i=1,2,\dots,\frac{n}{K}$ \textbf{do}\\
\hspace*{20pt} $\mathbf{x}_i\leftarrow$ the $i$-th codeword of $\mathbf{C}_b$\\
\hspace*{20pt} $N_i\leftarrow$ number of coordinates $j$ for which $\mathbf{x}_i(j)>\mathbf{y}(j)$\\
\hspace*{20pt} \textbf{If} $\  N_i\leq e\  $ \textbf{then}\\
\hspace*{40pt} Set $\mathcal{X}\leftarrow\mathcal{X}\cup\{\mathbf{x}_i\}$\\
\hspace*{20pt} \textbf{End}\\
\hspace*{0pt} \textbf{End}\\

\textbf{Step 2:} \\
\hspace*{0pt} \textbf{For} $i=1,2,\dots,|\mathcal{X}|$ \textbf{do}\\
\hspace*{20pt} Set $\mathcal{S}_i\leftarrow\{\textnormal{the set of nonzero coordinates of } \mathbf{x}_i\}$\\
\hspace*{20pt} Set $\mathcal{S}'_i\leftarrow\{$subset of $\mathcal{S}_i$ with $|\mathcal{S}'_i|=2e+1$ s.t. $\forall k\in\mathcal{S}'_i$ and $\forall j\in\mathcal{S}_i\backslash\mathcal{S}'_i$, one has $\mathbf{y}(k)\leq\mathbf{y}(j)\}$\\
\hspace*{20pt} Initialize the multiset $\mathcal{B}'\leftarrow\varnothing$\\
\hspace*{20pt} \textbf{For} $j=1,2,\dots,|\mathcal{S}'_i|$ \textbf{do}\\
\hspace*{40pt} $\eta_u\leftarrow$ the upper threshold of the quantization bin of $\mathbf{y}(j)$\\
\hspace*{40pt} $\eta_l\leftarrow$ the lower threshold of the quantization bin of $\mathbf{y}(j)$\\
\hspace*{40pt} $\beta\leftarrow$ the integer $\eta_l\leq\beta<\eta_u$ that can be written as the sum of up to $d$ elements of $\mathcal{A}$\\
\hspace*{65pt} (use Proposition~\ref{lemma_ks})\\
\hspace*{40pt} Update the multiset $\mathcal{B}'\leftarrow\mathcal{B}'\cup\{\beta\}$\\
\hspace*{20pt} \textbf{End}\\
\hspace*{20pt} Set $\hat{\beta}_t\leftarrow$ the element of $\mathcal{B}'$ with at least $e+1$ repetitions\\
\hspace*{20pt} Set $\hat{\mathcal{A}}_{i,t}\leftarrow\{$the unique subset of $\mathcal{A}$ with the sum equal to $\hat{\beta}_t\}$ \\
\hspace*{20pt} {Set $\hat{\mathcal{D}}_i\leftarrow\{\textnormal{codewords of }\mathbf{C} \textnormal{ of the form } \mathbf{z}=\alpha\mathbf{x}_i,\  \forall \alpha\in\hat{\mathcal{A}}_{i,t}       \}$}\\
\hspace*{0pt} \textbf{End}\\

\textbf{Return} $\hat{\mathcal{D}}=\bigcup_i\hat{\mathcal{D}}_i$
\vspace{-8pt}
\begin{center}
\line(1,0){469}
\end{center}
\vspace{-20pt}
\end{table}

\begin{lemma}\label{lemma_prune}
Consider a SQ-separable code constructed using Thm.~\ref{const_SQLO} and let $\mathbf{y}$ be the vector of test results with at most $e$ erroneous entries. Fix any binary codeword $\mathbf{x}_i\in\mathcal{X}_{\mathcal{D}}$, and let $\mathcal{S}_i$ be the set of nonzero coordinates of $\mathbf{x}_i$. Also, let $\mathcal{S}'_i\subseteq\mathcal{S}_i$, with $|\mathcal{S}'_i|=2e+1$, be the set of coordinates such that for any fixed $ k\in\mathcal{S}'_i$, one has $\mathbf{y}(k)\leq\mathbf{y}(j)$, $\forall j\in\mathcal{S}_i\backslash\mathcal{S}'_i$. Then, there exists a set $\mathcal{S}''_i\subseteq\mathcal{S}'_i$ such that $|\mathcal{S}''_i|\geq e+1$, and $\forall j\in\mathcal{S}''_i$, one has $\mathbf{y}(j)=\mathbf{y}_{\!_\mathcal{D}}(j)=f_{\boldsymbol{\eta}}\left(\sum_{\alpha\in\mathcal{A}_{i,t}}\alpha\right)$; in this equation, $\mathcal{A}_{i,t}\subseteq\mathcal{A}$ denotes the set of coefficients corresponding to the defective codewords with the same support as $\mathbf{x}_i$\footnote{As an example, assume that $\mathbf{x}_i\in\mathcal{X}_{\mathcal{D}}$ and let $\{\alpha_{j_1}\mathbf{x}_i,\alpha_{j_2}\mathbf{x}_i,\alpha_{j_3}\mathbf{x}_i\}\in\mathcal{D}$ be the only codewords in $\mathbf{C}$ with the same support as $\mathbf{x}_i$ in $\mathcal{D}$. In this case, $\mathcal{A}_{i,t}=\{\alpha_{j_1},\alpha_{j_2},\alpha_{j_3}\}$.}.  
\end{lemma}
\begin{IEEEproof}
Let $\mathcal{R}_i$ be the maximal set of coordinates such that $\forall j\in\mathcal{R}_i$, $\mathbf{x}_i(j)=1$ and $\mathbf{x}(j)=0$ for all $\mathbf{x}\in\mathcal{X}_{\mathcal{D}}\backslash\{\mathbf{x}_i\}$. Since $\mathbf{x}_i$ is a codeword of $\mathbf{C}_b$ and since $|\mathcal{X}_{\mathcal{D}}|\leq d$, the disjunctness property implies that such a set exists and that $|\mathcal{R}_i|\geq 2e+1$; clearly, $\mathcal{R}_i\subseteq\mathcal{S}_i$. Let $\mathcal{A}_{i,t}$ be the set of coefficients used to generate the codewords in $\mathcal{D}$ with the same support as $\mathbf{x}_i$. For all $k\in\mathcal{R}_i$, one has $\sum_{\mathbf{z}\in\mathcal{D}}\mathbf{z}(k)=\sum_{\alpha\in\mathcal{A}_{i,t}}\alpha$, and $\forall j\in\mathcal{S}_i\backslash\mathcal{R}_i$, one has $\sum_{\mathbf{z}\in\mathcal{D}}\mathbf{z}(j)>\sum_{\alpha\in\mathcal{A}_{i,t}}\alpha$. Note that the strict inequality follows since $\mathcal{R}_i$ is a maximal set. 
Since all the sums of up to $d$ elements of $\mathcal{A}$ fall into different quantization bins, for any $k\in\mathcal{R}_i$ and for any $j\in\mathcal{S}_i\backslash\mathcal{R}_i$, one has
\begin{align}\nonumber
f_{\boldsymbol{\eta}}\left(\sum_{\mathbf{z}\in\mathcal{D}}\mathbf{z}(k)\right)<f_{\boldsymbol{\eta}}\left(\sum_{\mathbf{z}\in\mathcal{D}}\mathbf{z}(j)\right).
\end{align}
As a result, if there were no errors in $\mathbf{y}$, one would have $\mathcal{S}'_i\subseteq\mathcal{R}_i$. Each erroneous entry of $\mathbf{y}$ removes at most one coordinate of $\mathcal{R}_i$ from $\mathcal{S}'_i$. Since there are at most $e$ errors and $|\mathcal{S}'_i|=2e+1$, there exists a set of coordinates $\mathcal{S}''_i\subseteq\mathcal{S}'_i\cap\mathcal{R}_i$ with cardinality at least $e+1$ for which the corresponding entries of $\mathbf{y}$ are error-free. As a result, $\forall j\in\mathcal{S}''_i$ one has $\mathbf{y}(j)=\mathbf{y}_{\!_\mathcal{D}}(j)=f_{\boldsymbol{\eta}}\left(\sum_{\alpha\in\mathcal{A}_{i,t}}\alpha\right)$.

\end{IEEEproof}

\begin{prop}\label{lemma_ks}
Given a SQLO$_s(\boldsymbol{\eta},d)$ sequence $\mathcal{A}$ and a fixed integer $\beta$, one can identify whether $\beta$ can be written as a sum of up to $d$ elements of $\mathcal{A}$ with an algorithm of computational complexity $O(K)$, where $K=|\mathcal{A}|$. Given that the answer to this question is positive, one can identify the elements of $\mathcal{A}$ which sum up to $\beta$ with computational complexity $O(K)$. 
\end{prop}
\begin{IEEEproof}
This problem is known as the knapsack-solvability problem~\cite{PHS03}. From the second property of Proposition~\ref{prop_1}, $\forall i:1\leq i\leq K$ and $\forall \mathcal{A}_1\subseteq\{\alpha_1,\alpha_2,\dots,\alpha_{i-1}\}$ such that $|\mathcal{A}_1|\leq d$, one has $\alpha_i\succ_{\boldsymbol{\eta}}\sum_{\alpha_j\in\mathcal{A}_1}\alpha_j$, which also implies that $\alpha_i>\sum_{\alpha_j\in\mathcal{A}_1}\alpha_j$. 

To find the answer to the query with linear computational complexity, we perform a standard knapsack recursion~\cite{PHS03}. First, we initialize the procedure by setting $\beta'\leftarrow\beta$ and $\hat{\mathcal{A}}_\beta\leftarrow\varnothing$. Then, in the $i$-th iteration, we compare the value of $\beta'$ with the $i$-th largest element of $\mathcal{A}$, $\alpha_{K-i+1}$. If $\beta'\geq\alpha_{K-i+1}$, then we update $\beta'\leftarrow\beta'-\alpha_{K-i+1}$ and $\hat{\mathcal{A}}_\beta\leftarrow\hat{\mathcal{A}}_\beta\cup\{\alpha_{K-i+1}\}$; otherwise, we go to the next iteration. The procedure stops with a negative answer to the first query if $|\hat{\mathcal{A}}_\beta|>d$ or if $\beta'>0$ and no element in $\mathcal{A}$ is left that is smaller than or equal to $\beta'$. Otherwise, the procedure stops when $\beta'=0$ with a positive answer to the first query, and $\hat{\mathcal{A}}_\beta$ corresponds to the elements of $\mathcal{A}$ that sum up to $\beta$. Note that this procedure is based on the superincreasing property of a SQLO$_s$ sequence, which implies that the largest element of $\mathcal{A}$ that does not exceed $\beta'$ must be present in the sum.
\end{IEEEproof}

The previous proposition and lemma provide the core of the second step of Algorithm~2. The idea is that for each $\mathbf{x}_i\in\mathcal{X}_{\mathcal{D}}$, we use Lemma~\ref{lemma_prune} to find $\mathcal{S}'_i$. The majority of elements $\mathbf{y}(j)$, $j\in\mathcal{S}'_i$, correctly correspond to the bin in which $\beta_t=\sum_{\alpha\in\mathcal{A}_{i,t}}\alpha$ is located. Each correctly identified bin contains a finite number of integers, one of which is the true value of $\beta_t$. As a result, by testing each such integer, we can determine whether it can be written as the sum of up to $d$ elements of $\mathcal{A}$ or not using the algorithm in Proposition~\ref{lemma_ks}. The integer for which the answer to this query is positive is equal to $\beta_t$, which can then be used to identify the elements of $\mathcal{A}_{i,t}$.

\begin{theorem}
The Dec-SQLO$_s$ algorithm is capable of identifying up to $d$ defectives in the presence of at most $e$ errors in the syndrome of defectives.
\end{theorem}
\begin{IEEEproof}
Since the first step of this algorithm is identical to the first step of the Dec-QBh algorithm, it follows that $\mathcal{X}=\mathcal{X}_\mathcal{D}$. Therefore, we only need to show that Step 2 recovers $\mathcal{D}$ given $\mathcal{X}_{\mathcal{D}}$. 

Fix a binary vector $\mathbf{x}_i\in\mathcal{X}_{\mathcal{D}}$. Fix a coordinate $j\in\mathcal{S}'_i$, and let $\eta_l$ and $\eta_u$ be the lower and upper thresholds of the quantization bin corresponding to $\mathbf{y}(j)$. Since all the sums of up to $d$ elements of $\mathcal{A}$ fall into different bins, there exists exactly one subset sum $\beta$ in $[\eta_l,\eta_u)$ that corresponds to the sum of up to $d$ elements of $\mathcal{A}$. As a result, one can test all the $(\eta_u-\eta_l)$ integers in this bin using Proposition~\ref{lemma_ks} to find the unique value of $\beta$ that can be written as sum of up to $d$ elements of $\mathcal{A}$. 
On the other hand, as was shown in Lemma~\ref{lemma_prune}, there exists a set $\mathcal{S}''_i\subseteq\mathcal{S}'_i$ such that $|\mathcal{S}''_i|\geq e+1$, and consequently $\forall j\in\mathcal{S}''_i$ one has $\mathbf{y}(j)=f_{\boldsymbol{\eta}}\left(\sum_{\alpha\in\mathcal{A}_{i,t}}\alpha\right)=f_{\boldsymbol{\eta}}\left(\beta_t\right)$. As a result, the element $\beta$ in the multiset $\mathcal{B}'$ with multiplicity at least $e+1$ corresponds to $\beta_t$, or in other words $\hat{\beta}_t=\beta_t$. This implies that $\hat{\mathcal{A}}_{i,t}=\mathcal{A}_{i,t}$, and consequently, $\hat{\mathcal{D}}=\mathcal{D}$.
\end{IEEEproof}

\begin{remark}
The computational complexity of the Dec-SQLO$_s$ algorithm is equal to $O(\frac{mn}{K}+dm\log m+deg_{\max}K)$, where $g_{\max}=\max_{i=1,2,\dots,Q}\left(\eta_i-\eta_{i-1}\right)$ is the largest gap between the consecutive thresholds. The computational complexity of Step 1 is $O(\frac{mn}{K})$. On the other hand, sorting the elements of $\mathcal{S}_i$ to find $\mathcal{S}'_i$ requires $O(dm\log m)$ computations. One can identify the elements of $\mathcal{A}$ that sum up to a fixed integer in linear time., i.e. using $O(K)$ computational steps. As a result, the algorithm for finding $\beta$ in each iteration has complexity $O(eg_{\max}K)$. Hence, finding $\hat{\mathcal{A}}_{i,t}$ requires $O(deg_{\max}K)$ computational steps.
\end{remark}

\section{SQ-separable codes using SQLO$_l$ sequences}  \label{sec:sqlol}
The SQLO$_s$ sequences introduced in the previous section resolves the problem of exponential growth of decoding computational complexity with respect to $K$. However, due to the superincreasing property of these sequences (the second property in Prop.~\ref{prop_1}) the multipliers $\alpha_K$ tend to grow rapidly as a function of $K$. In order to overcome this issue while preserving efficient decoding, we introduce a new family of integer sequences, termed SQLO$_l$ sequences. 

\begin{defin}[\textbf{SQLO$_l(\boldsymbol{\eta},h)$ sequences}]\label{SQLO2}
Given a set of thresholds $\boldsymbol{\eta}$, a sequence of positive integers $\mathcal{A}=\{\alpha_1,\alpha_2,\dots,\alpha_K\}$ is a SQLO$_l(\boldsymbol{\eta},h)$ sequence if
\begin{enumerate}
\item $\alpha_K\succ_{\boldsymbol{\eta}}\alpha_{K-1}\succ_{\boldsymbol{\eta}}\dots\alpha_2\succ_{\boldsymbol{\eta}}\alpha_1\succ_{\boldsymbol{\eta}}0$ (i.e., all elements of $\mathcal{A}$ lie in different quantization bins).
\item For any two subsets $\mathcal{A}_1\subseteq\mathcal{A}$ and $\mathcal{A}_2\subseteq\mathcal{A}$ such that $|\mathcal{A}_1|<|\mathcal{A}_2|\leq h$, one has $\sum_{\alpha_i\in\mathcal{A}_2}\alpha_i\succ_{\boldsymbol{\eta}}\sum_{\alpha_i\in\mathcal{A}_1}\alpha_i$ (i.e., subsets of different cardinality are ordered based on the number of their members).
\item For any two distinct subsets $\mathcal{A}_1=\{\alpha'_1,\alpha'_2,\dots,\alpha_s'\}$ and $\mathcal{A}_2=\{\alpha''_1,\alpha''_2,\dots,\alpha''_s\}$ with elements listed in an increasing order such that $|\mathcal{A}_1|=|\mathcal{A}_2|=s\leq h$, one has $\sum_{\alpha''_i\in\mathcal{A}_2}\alpha''_i\succ_{\boldsymbol{\eta}}\sum_{\alpha'_i\in\mathcal{A}_1}\alpha'_i$ if there exists $r:1\leq r\leq s$ such that $\forall i:1\leq i<r$, $\alpha'_i=\alpha''_i$ and $\alpha''_r\succ_{\boldsymbol{\eta}}\alpha'_r$ (i.e., two subsets with the same cardinality are lexicographically ordered).
\end{enumerate}
\end{defin}

The following example illustrates how SQLO$_l$ properties may lead to denser sequences compared to SQLO$_s$ sequences.

\begin{example}
As an example, consider the set of thresholds $\boldsymbol{\eta}=[0,2,5,6,10,11,15,18]^T$ and let $h=2$ and $K=3$. The sequence $\mathcal{A}_1=\{2,5,10\}$ is a SQLO$_s(\boldsymbol{\eta},2)$ sequence that has the smallest value for $\alpha_3$, i.e. $\alpha_3=10$. On the other hand, the sequence $\mathcal{A}_2=\{4,5,6\}$ is a SQLO$_l(\boldsymbol{\eta},2)$ sequence that has the smallest value for $\alpha_3$, i.e. $\alpha_3=6$. 
\end{example}

The SQLO$_l$ properties impose a partial order on the subsets of the sequence. For example, if $K=3$ and $h\geq K$, these properties translate into $\alpha_3+\alpha_2+\alpha_1\succ_{\boldsymbol{\eta}}\alpha_3+\alpha_2\succ_{\boldsymbol{\eta}}\alpha_3+\alpha_1\succ_{\boldsymbol{\eta}}\alpha_2+\alpha_1\succ_{\boldsymbol{\eta}}\alpha_3\succ_{\boldsymbol{\eta}}\alpha_2\succ_{\boldsymbol{\eta}}\alpha_1\succ_{\boldsymbol{\eta}}0$. Similarly to the case of SQLO$_s$ sequences, it is not difficult to see that any SQLO$_l(\boldsymbol{\eta},h)$ sequence is also a quantized $B_h$ sequence; however, the converse is not necessarily true. As a result, the following theorem holds.

\begin{theorem}\label{const_SQLO2}
Fix a binary $d$-disjunct code matrix $\mathbf{C}_b$ of dimensions $m_b\times n_b$, capable of correcting up to $e$ errors. Let $\mathcal{A}=\{\alpha_1,\alpha_2,\dots,\alpha_K\}$ be a SQLO$_l(\boldsymbol{\eta},d)$ sequence. Form a matrix $\mathbf{C}$ of length $m=m_b$ and size $n=Kn_b$ by concatenating $K$ matrices $\mathbf{C}_i=\alpha_i\mathbf{C}_b$, $1\leq i\leq K$ horizontally. The constructed code is a $[q;Q;\boldsymbol{\eta};(1\!:\!d);e]$-SQ-separable code with $q=\alpha_K+1$.\end{theorem} 

\begin{IEEEproof}
Since any SQLO$_l(\boldsymbol{\eta},d)$ sequence is a quantized $B_d$ sequence, the proof follows directly from Thm.~\ref{const_Bh}. 
\end{IEEEproof}

\subsection{Fundamental limits and construction of SQLO$_l$ sequences}


In~\cite{ANS90}, two types of lexicographically ordered sequences were defined that are closely related to the SQLO$_l$ sequences. For simplicity, we call these sequences ``lex($h$)'' and ``strong-lex($h$)'' and we provide their definition for completeness.

\begin{defin}\label{def_lex}
A sequence of positive integers $\mathcal{B}=\{\beta_1,\beta_2,\dots\}$ is a lex($h$) sequence, if for any two distinct subsets $\mathcal{B}_1=\{\beta'_1,\beta'_2,\dots,\beta_h'\}$ and $\mathcal{B}_2=\{\beta''_1,\beta''_2,\dots,\beta''_h\}$ with elements listed in an increasing order, one has $\sum_{\beta''_i\in\mathcal{B}_2}\beta''_i>\sum_{\beta'_i\in\mathcal{B}_1}\beta'_i$ if there exists an integer $r$, $1\leq r\leq h$, such that $\forall i$, $1\leq i<r$, $\beta'_i=\beta''_i$ and $\beta''_r>\beta'_r$.
\end{defin}

\begin{defin}\label{def_strong_lex}
A sequence of positive integers $\mathcal{B}=\{\beta_1,\beta_2,\dots\}$ is a strong-lex($h$) sequence, if it is a lex($s$) sequence $\forall s\leq h$; in addition, for any two subsets $\mathcal{B}_1\subseteq\mathcal{B}$ and $\mathcal{B}_2\subseteq\mathcal{B}$ such that $|\mathcal{B}_1|<|\mathcal{B}_2|\leq h$, one has $\sum_{\beta_i\in\mathcal{B}_2}\beta_i>\sum_{\beta_i\in\mathcal{B}_1}\beta_i$.

\end{defin}

The strong-lex($h$) sequences can be used to construct SQLO$_l$ sequences as shown in the next proposition.

\begin{prop}\label{SQLO2_const}
Consider a SQGT model with thresholds $\boldsymbol{\eta}=[0,\eta_1,\eta_2,\dots,\eta_Q]^T$; $\forall s:1\leq s\leq Q$, let $g_s=\max_{i:1\leq i\leq s}\eta_i-\eta_{i-1}$ be the largest gap of the first $s$ thresholds. Let $\mathcal{B}=\{\beta_1<\beta_2<\dots\}$ be a strong-lex($h$) sequence. For a fixed $s$, $2\leq s\leq Q$, let $K_s$ be the largest positive integer that satisfies $\eta_s> g_s\sum_{i=\max\{1,K_s-h\}}^{K_s} \beta_i$. Then all the sequences of the form $\mathcal{A}_s = \left\{g_s\:\beta_1, g_s\:\beta_2, \dots, g_s\:\beta_{K_s}\right\}$ are SQLO$_l(\boldsymbol{\eta},h)$ sequences. 
\end{prop}
\begin{IEEEproof}
The proof of this proposition follows along the same lines as the proof of Thms.~\ref{qbh_const} and \ref{SQLO1_const}, and is hence omitted. 
\end{IEEEproof}

As we demonstrated through a simple example earlier, the SQLO$_l$ properties may result in denser sequences compared to SQLO$_s$ properties. However, a SQLO$_l(\boldsymbol{\eta},h)$ sequence constructed from strong-lex($h$) sequences according to Proposition~\ref{SQLO2_const} does not improve the bound $\alpha_K=O_g\left(\gamma^{K}\right)$ derived in Lemma~\ref{prop_recur}. This can be shown as follows. We define an optimal lex($h$) sequence as a lex($h$) sequence $\mathcal{B}=\{\beta_1,\beta_2,\dots,\beta_K\}$ with the smallest possible value of $\beta_K$. In~\cite[Thm. 1]{ANS90}, it was proven that the largest element of an optimal lex($h$) sequence satisfies $\beta_K=O\left(\gamma^K\right)$, where $\gamma$ is the largest root of $x^{h+1}-2x^h+1=0$\footnote{Note that there exists a typo in the statement of~\cite[Thm. 1]{ANS90}, in which $\gamma$ is defined as the largest root of $x^{h+1}-x^h+1=0$. However, it is evident from the proof of the theorem that $\gamma$ is in fact the largest root of $x^{h+1}-2x^h+1=0$.}. Since any strong-lex($h$) sequence needs to also satisfy the lex($h$) property, one can conclude that a SQLO$_l(\boldsymbol{\eta},h)$ sequence constructed from strong-lex($h$) sequences according to Proposition~\ref{SQLO2_const} cannot improve the bound $\alpha_K=O_g\left(\gamma^{K}\right)$.

\subsection{Decoding algorithm for SQGT codes constructed using SQLO$_l$ sequences}
Next, we describe the Dec-SQLO$_l$ algorithm, the decoding procedure for codes based on SQLO$_l$ sequences. This algorithm resembles the Dec-SQLO$_s$ algorithm, and similar intuition also applies as follows. In the first step, one identifies $\mathcal{X}=\mathcal{X}_{\mathcal{D}}$, the set of binary codewords corresponding to the support of the codewords in $\mathcal{D}$. To complete the decoding, one needs to identify the set of elements $\mathcal{A}_{i,t}\subseteq\mathcal{A}$ which are used to form the codewords in $\mathcal{D}$ from the binary codeword $\mathbf{x}_i$, $\forall\mathbf{x}_i\in\mathcal{X}_{\mathcal{D}}$. 

Since in the proof of Lemma~\ref{lemma_prune} we only used the quantized $B_d$ property, this lemma also holds for codes constructed using Thm.~\ref{const_SQLO2}. As a result, using the notation defined in Lemma~\ref{lemma_prune}, for any binary codeword $\mathbf{x}_i\in\mathcal{X}_{\mathcal{D}}$ there exists at least $e+1$ elements of $\mathcal{S}'_i$, denoted by $\mathcal{S}''_i$, such that $\forall j\in\mathcal{S}''_i$ one has $\mathbf{y}(j)=f_{\boldsymbol{\eta}}\left(\sum_{\alpha\in\mathcal{A}_{i,t}}\alpha\right)$. This implies that the majority of elements $\mathbf{y}(j)$, $j\in\mathcal{S}'_i$, correctly correspond to the bin in which $\beta_t=\sum_{\alpha\in\mathcal{A}_{i,t}}\alpha$ is located. Each correctly identified bin contains a limited number of integers, one of which is the true value of $\beta_t$. As a result, by testing each such integer, we can determine whether it can be written as the sum of up to $d$ elements of $\mathcal{A}$ or not. The integer for which the answer to this query is positive is equal to $\beta_t$, which can then be used to identify the elements of $\mathcal{A}_{i,t}$. 

As a result, given a SQLO$_l$ sequence $\mathcal{A}$ and an integer $\beta$, the main issue is to efficiently determine whether $\beta$ can be written as the sum of up to $d$ elements of $\mathcal{A}$; and if so, what those elements are. In Lemma~\ref{lemma_ks2}, an algorithm with computational complexity of $O(K)$ is described that can perform this task.

\begin{lemma}\label{lemma_ks2}
Given a SQLO$_l(\boldsymbol{\eta},d)$ sequence $\mathcal{A}$ and a fixed integer $\beta$, it is possible to identify whether $\beta$ can be written as a sum of up to $d$ elements of $\mathcal{A}$ with complexity $O(K)$, where $K=|\mathcal{A}|$. Given that the answer to this question is positive, one can identify these elements of $\mathcal{A}$ with complexity $O(K)$. 
\end{lemma}
\begin{IEEEproof}
Suppose $\beta$ can be written as the sum of $s\leq d$ elements of $\mathcal{A}$, and let $\mathcal{A}_t\subseteq\mathcal{A}$ be the subset such that $\sum_{\alpha\in\mathcal{A}_t}\alpha=\beta$. The value of $s=|\mathcal{A}_t|$ can be easily determined as follows. First, we form the set $\Gamma=\{\gamma_1,\gamma_2,\dots,\gamma_K\}$, where $\gamma_i=\sum_{j=1}^i\alpha_i$, $1\leq i\leq K$. As a consequence of the second property in Def.~\ref{SQLO2}, for any $1\leq i\leq K$, $\gamma_i$ is larger than all $j$-subsets of $\mathcal{A}$ for $j<i$. On the other hand, due to the third property in Def.~\ref{SQLO2}, $\gamma_i$ is smaller than all $j$-subsets of $\mathcal{A}$ for $j\geq i$. Consequently, one can determine $s$ using $s=\min\{i:\beta<\gamma_i\}-1$.

Given the value of $s$, we can determine the elements of $\mathcal{A}_t$ successively using $K$ iterations. First, we initialize the procedure by setting $s'\leftarrow s$, $\beta'\leftarrow\beta$, and $\mathcal{A}'\leftarrow\varnothing$. In the $i$-th iteration, $1\leq i\leq K$, we determine whether $\alpha_i\in\mathcal{A}_t$ or not. At the beginning of the $i$-th iteration, $\mathcal{A}'$ equals $\mathcal{A}_t\cap\{\alpha_1,\alpha_2,\dots,\alpha_{i-1}\}$, and $s'$ is equal to the number of remaining unidentified elements of $\mathcal{A}_t$, i.e. $s'=|\mathcal{A}_t\backslash\mathcal{A}'|$. In addition, $\beta'$ is equal to the sum of $s'$ elements in $\mathcal{A}_t\backslash\mathcal{A}'$. To determine whether $\alpha_i$ is in $\mathcal{A}_t$, we use the following rule: if $\beta'<\alpha_{i+1}+\alpha_{i+2}+\dots+\alpha_{i+s'}$, then $\alpha_i\in\mathcal{A}_t$; the reason is that the sum of any $s'$ elements of $
\{\alpha_i,\alpha_{i+1},\dots,\alpha_{K}\}$ that does not include $\alpha_i$ is at least as large as $\alpha_{i+1}+\alpha_{i+2}+\dots+\alpha_{i+s'}$. Therefore, if $\beta'<\alpha_{i+1}+\alpha_{i+2}+\dots+\alpha_{i+s'}$, then $\alpha_i$ must be in $\mathcal{A}_t$. Given that this condition is satisfied, we update $\mathcal{A}'\leftarrow\mathcal{A}'\cup\{\alpha_i\}$, $\beta'\leftarrow\beta'-\alpha_i$, and $s'\leftarrow s'-1$. Otherwise, we go to the next iteration. The algorithm stops after $K$ iterations. At the end, if $s'=0$ and $|\mathcal{A}'|\leq d$, the answer to the first query is positive and $\mathcal{A}'=\mathcal{A}_t$. Otherwise the answer to the query is negative. 
\end{IEEEproof}

The decoding algorithm for codes constructed using Thm.~\ref{const_SQLO2} is described in Algorithm 3. Note that the main difference between this algorithm and Algorithm 2 is that in Step 2, we use Lemma~\ref{lemma_ks2} instead of Proposition~\ref{lemma_ks}.

\begin{table}
\begin{center}
\line(1,0){469}
\end{center}
\vspace{-8pt}
\textbf{Algorithm 3: Dec-SQLO$_l$} 
\vspace{-10pt}
\begin{center}
\line(1,0){469}
\end{center}
\textbf{Input:} $\mathbf{y}\in[Q]^{m}$, $\mathbf{C}_b\in[2]^{m\times \frac{n}{K}}$, $\boldsymbol{\eta}$, $\mathcal{A}$, $e\geq0$ \\
\textbf{Output:} $\hat{\mathcal{D}}$\\

\textbf{Step 1:} Initialize $\mathcal{X}\leftarrow\varnothing$ and $\hat{\mathcal{D}}\leftarrow\varnothing$\\
\hspace*{0pt} \textbf{For} $i=1,2,\dots,\frac{n}{K}$ \textbf{do}\\
\hspace*{20pt} $\mathbf{x}_i\leftarrow$ the $i$-th codeword of $\mathbf{C}_b$\\
\hspace*{20pt} $N_i\leftarrow$ number of coordinates $j$ for which $\mathbf{x}_i(j)>\mathbf{y}(j)$\\
\hspace*{20pt} \textbf{If} $\  N_i\leq e\  $ \textbf{then}\\
\hspace*{40pt} Set $\mathcal{X}\leftarrow\mathcal{X}\cup\{\mathbf{x}_i\}$\\
\hspace*{20pt} \textbf{End}\\
\hspace*{0pt} \textbf{End}\\

\textbf{Step 2:} \\
\hspace*{0pt} \textbf{For} $i=1,2,\dots,|\mathcal{X}|$ \textbf{do}\\
\hspace*{20pt} Set $\mathcal{S}_i\leftarrow\{\textnormal{the set of nonzero coordinates of } \mathbf{x}_i\}$\\
\hspace*{20pt} Set $\mathcal{S}'_i\leftarrow\{$subset of $\mathcal{S}_i$ with $|\mathcal{S}'_i|=2e+1$ s.t. $\forall k\in\mathcal{S}'_i$ and $\forall j\in\mathcal{S}_i\backslash\mathcal{S}'_i$, one has $\mathbf{y}(k)\leq\mathbf{y}(j)\}$\\
\hspace*{20pt} Initialize the multiset $\mathcal{B}'\leftarrow\varnothing$\\
\hspace*{20pt} \textbf{For} $j=1,2,\dots,|\mathcal{S}'_i|$ \textbf{do}\\
\hspace*{40pt} $\eta_u\leftarrow$ the upper threshold of the quantization bin for $\mathbf{y}(j)$\\
\hspace*{40pt} $\eta_l\leftarrow$ the lower threshold of the quantization bin for $\mathbf{y}(j)$\\
\hspace*{40pt} $\beta\leftarrow$ the unique integer $\eta_l\leq\beta<\eta_u$ that can be written as the sum of up to $d$ elements\\
 \hspace*{65pt} of $\mathcal{A}$ (use Lemma~\ref{lemma_ks2})\\
\hspace*{40pt} Update the multiset $\mathcal{B}'\leftarrow\mathcal{B}'\cup\{\beta\}$\\
\hspace*{20pt} \textbf{End}\\
\hspace*{20pt} Set $\hat{\beta}_t\leftarrow$ the element of $\mathcal{B}'$ with at least $e+1$ repetitions\\
\hspace*{20pt} Set $\hat{\mathcal{A}}_{i,t}\leftarrow\{$the unique subset of $\mathcal{A}$ with the sum equal to $\hat{\beta}_t\}$ \\
\hspace*{20pt} {Set $\hat{\mathcal{D}}_i\leftarrow\{\textnormal{codewords of }\mathbf{C} \textnormal{ of the form } \mathbf{z}=\alpha\mathbf{x}_i,\  \forall \alpha\in\hat{\mathcal{A}}_{i,t}       \}$}\\
\hspace*{0pt} \textbf{End}\\

\textbf{Return} $\hat{\mathcal{D}}=\bigcup_i\hat{\mathcal{D}}_i$
\vspace{-8pt}
\begin{center}
\line(1,0){469}
\end{center}
\vspace{-20pt}
\end{table}
\section*{Acknowledgment}
This work was supported in part by NSF grants CCF 0809895, CCF 1218764 and the Emerging Frontiers for Science of Information Center, CCF 0939370. The authors would like to thank Farzad Farnoud (Hassanzadeh), Han Mao Kiah, and Gregory Puleo for useful discussions.


\end{document}